\PassOptionsToPackage{table}{xcolor}
\documentclass{lmcs}
\pdfoutput=1
\usepackage[utf8]{inputenc}

\usepackage{lastpage}
\lmcsdoi{22}{2}{26}
\lmcsheading{}{\pageref{LastPage}}{}{}%
{Dec.~11,~2023}{Jun.~11,~2026}{}


\usepackage{hyperref}

\usepackage{tikz}

\newcommand{\tw}{\ensuremath\textrm{tw}}
\newcommand{\pw}{\ensuremath\textrm{pw}}
\newcommand{\td}{\ensuremath\textrm{td}}

\newcommand{\tow}{\ensuremath\textrm{tow}}

\begin{document}

\title{First Order Logic on Pathwidth Revisited Again} 

\thanks{This work was partially supported by the ANR project ANR-21-CE48-0022 (S-EX-AP-PE-AL).}

\author[M.~Lampis]{Michael Lampis\lmcsorcid{0000-0002-5791-0887}}

\address{Université Paris-Dauphine, PSL University, CNRS, LAMSADE, 75016, Paris, France}
\email{michail.lampis@lamsade.dauphine.fr}

\begin{abstract}

Courcelle's celebrated theorem states that all MSO-expressible properties can
be decided in linear time on graphs of bounded treewidth. Unfortunately, the
hidden constant implied by this theorem is a tower of exponentials whose height
increases with each quantifier alternation in the formula. More devastatingly,
this cannot be improved, under standard assumptions, even if we consider the
much more restricted problem of deciding FO-expressible properties on trees.

In this paper we revisit this well-studied topic and identify a natural special
case where the dependence of Courcelle's theorem can, in fact, be improved.
Specifically, we show that all FO-expressible properties can be decided with an
elementary dependence on the input formula if the input graph has bounded
pathwidth (rather than treewidth).  This is a rare example of treewidth and
pathwidth having different complexity behaviors. Our result is also in sharp
contrast with MSO logic on graphs of bounded pathwidth, where it is known that
the dependence has to be non-elementary, under standard assumptions.  Our work
builds upon, and generalizes, a corresponding meta-theorem by Gajarsk{\'{y}}
and Hlin{\v{e}}n{\'{y}} for the more restricted class of graphs of bounded
tree-depth.

\end{abstract}

\maketitle

\section{Introduction}

Algorithmic meta-theorems are general statements of the form ``all problems in
 certain class are tractable on a particular class of inputs''. Probably the
most famous and celebrated result of this type is Courcelle's theorem
\cite{Courcelle90}, which states that all graph properties expressible in
Monadic Second Order (MSO) logic are solvable in linear time on graphs of
bounded treewidth. This result has proved to be of immense importance to
parameterized complexity theory, because a vast collection of natural NP-hard
problems can be expressed in MSO logic (and its variations that allow
optimization objectives \cite{ArnborgLS91}) and because treewidth is the most
well-studied structural graph parameter. Thanks to Courcelle's theorem, we
immediately obtain that all such problems are fixed-parameter tractable (FPT)
parameterized by treewidth.

Despite its great success, Courcelle's theorem suffers from a significant
weakness: the algorithm it guarantees has a running time that is astronomical
for most problems.  Indeed, a careful reading of the theorem shows that the
running time increases as a tower of exponentials whose height is equal to the
number of quantifier alternations of the input MSO formula.  Hence, even though
Courcelle's theorem shows that any MSO formula $\phi$ can be decided on
$n$-vertex graphs of treewidth $\tw$ in time $f(\phi,\tw)n$, the function $f$
is \emph{non-elementary}, that is, it cannot be bounded from above by any tower
of exponentials of fixed height.

One could hope that this terrible dependence on $\phi$ is an artifact of
Courcelle's proof technique. Unfortunately, it was shown in a very influential
work by Frick and Grohe \cite{FrickG04} that this non-elementary dependence on
the number of quantifiers of $\phi$ is best possible (under standard
assumptions), even if one considers the severely restricted special case of
model-checking First Order (FO) logic on trees. Recall that FO logic is a basic
logic formalism that allows us to express graph properties using quantification
over the vertices of the graph, while MSO logic also allows quantification over
sets of vertices.  Since FO logic is trivially a subset of MSO logic and trees
have treewidth $1$, this result established that Courcelle's theorem is
essentially best possible.

Frick and Grohe's lower bound thus provided the motivation for the search for
subclasses of bounded-treewidth graphs where avoiding the non-elementary
dependence on $\phi$ may be possible. The obvious next place to look was
naturally, pathwidth, which is the most well-known restriction (and close
cousin) of treewidth. Unfortunately, Frick and Grohe's paper provided a
negative result for MSO model checking also for this parameter. More precisely,
they showed that MSO model checking on strings with a total order relation has
a non-elementary dependence on the formula (unless P=NP), but such structures
can easily be embedded into caterpillars (which are graphs of pathwidth $1$) if
one allows quantification over sets.  Notice, however, that this does not
settle the complexity of FO logic for graphs of counstant pathwidth, as it is
not clear how one could implement the total ordering relation of a string
without access to set quantifiers (we expand on this question further below).

On the positive side, Frick and Grohe's lower bounds motivated the discovery of
several meta-theorems with elementary dependence on the formula for other, more
restricted variations of treewidth (we review some such results below). Of all
these results, the one that is ``closest'' to treewidth is the theorem of
Gajarsk{\'{y}} and Hlin{\v{e}}n{\'{y}} \cite{Gajarsky15}, which states that on
graphs of constant tree-depth, MSO (and hence FO) model checking has elementary
dependence on the input formula. It is known that for all $n$-vertex graphs $G$
we have $\tw(G) \le \pw(G) \le \td(G) \le \tw(G)\log n$, where $\tw, \pw, \td$
denote the treewidth, pathwidth and tree-depth.  In a sense, this positive
result seemed to go as far as one could possibly go towards emulating
treewidth, while retaining the elementary dependence on the formula and
avoiding the lower bound of Frick and Grohe.  This state of the art is
summarized in Table~\ref{tab:state}.

\begin{table}

\begin{tabular}{l|l|l} 
\hline
Parameter & FO & MSO \\
\hline
Treewidth & \cellcolor{red!25}Non-elementary on Trees \cite{FrickG04} & \cellcolor{red!25}Non-elementary on Trees \cite{FrickG04} \\
Pathwidth & \cellcolor{green!55}Elementary (Theorem~\ref{thm:elem}) & \cellcolor{red!25}Non-elementary on Caterpillars \cite{FrickG04} \\
Tree-depth & \cellcolor{green!25}Elementary \cite{Gajarsky15}& \cellcolor{green!25}Elementary \cite{Gajarsky15} \\
\hline

\end{tabular}

\caption{Summary of the state of the art for FO and MSO model checking on
graphs of bounded treewidth, pathwidth, and tree-depth. Elementary (green
cells) indicates that there is an algorithm which, when the corresponding width
is bounded by an absolute constant, decides any formula $\phi$ in time
$f(\phi)n^{O(1)}$, where $f$ is a function that can be bounded above by a
finite tower of exponentials. For the remaining cases, this is known to be
impossible, under standard assumptions, hence it is inevitable to have an
$f(\phi)$ that is a tower of exponentials whose height increases with
$\phi$.}\label{tab:state}

\end{table}

\paragraph*{Our result:} In this paper we revisit this well-studied topic
and address the one remaining case of Table~\ref{tab:state} where it is still
unknown whether it is possible to obtain an elementary dependence on the
formula for model checking. We answer this question positively, showing that if
we restrict ourselves to graphs of pathwidth $p$, where $p$ is an absolute
constant, then FO formulas with $q$ quantifiers can be decided in time
$f(q)n^{O(1)}$, where $f$ is an elementary function of $q$. More precisely, the
function $f$ is at most a tower of exponentials of height $O(p)$. In other
words, our result trades the non-elementary dependence on $q$, which is inherent
in Courcelle's theorem, with a non-elementary dependence on $p$. Though this
may seem disappointing at first, it is known that this is the best one could
have hoped for. In fact, the meta-theorem of \cite{Gajarsky15} also has this
behavior (its parameter dependence is a tower of exponentials whose height
increases with the tree-depth), and it was shown in \cite{Lampis14} that this
is best possible (under standard assumptions). Since pathwidth is a more
general parameter, we cannot evade this lower bound and our algorithm needs to
have a non-elementary dependence on pathwidth if its dependence on the formula
is elementary.

The result we obtain is, therefore, in a sense best possible and fills a
natural gap in our knowledge regarding FO model checking for a well-studied
graph width. Beyond filling this gap, the fact that we are able to give a
positive answer to this question and obtain an algorithm with ``good''
dependence on the formula is interesting, and perhaps even rather striking, for
several reasons.  First, in many cases in this domain, it is impossible to
obtain an elementary dependence on $q$, no matter how much we are willing to
sacrifice on our dependence on the graph width, as demonstrated by the fact
that the lower bounds of Table~\ref{tab:state} apply for classes with the smallest
possible width (trees and caterpillars). Second, even though FO logic seems
much weaker than MSO logic in general, the complexities of model checking the
two logics seem to be similar (that is, at most one level of exponentiation
apart) for most parameters (we review some further examples below). Indeed, a
main contribution of \cite{Gajarsky15} was to prove that for graphs of bounded
tree-depth, the two logics are actually equivalent. It is therefore somewhat
unusual (for this context) that for pathwidth FO logic has quite different
complexity from MSO logic.  Third, even though treewidth and pathwidth are
arguably the two most well-studied graph widths in parameterized complexity, by
and large the complexities of the vast majority of problems are the same for
both parameters in the sense that problems solvable in time $f(\pw)n^{O(1)}$
are typically also solvable in time $f(\tw)n^{O(1)}$ for the same function $f$
(for more information on this, see \cite{BelmonteKLMO22} which only recently
discovered the first example of a natural problem separating the two
parameters). It is therefore remarkable that the complexity of FO model
checking is so different for pathwidth and treewidth.

Finally, one aspect of our result that makes it more surprising is that it does
not seem to generalize to dense graphs. Meta-theorems that give a
non-elementary dependence on the formula by using a restriction of treewidth,
generally have a dense graph analogue, using a restriction of clique-width (the
dense graph analogue of treewidth). Indeed, this is the case for vertex cover
\cite{Lampis12} (neighborhood diversity \cite{Lampis12}, twin cover
\cite{Ganian15}) but also for tree-depth (shrub-depth \cite{GanianHNOM19}).
One may have expected something similar to hold in our case.  However, the
natural dense analogue of pathwidth is linear clique-width and it is already
known that FO logic has a non-elementary dependence on threshold graphs
\cite{Lampis14}.  Since threshold graphs have linear clique-width $2$, we
cannot hope to extend our result to this parameter and it appears that the
positive result of this paper is an isolated island of ``tractability''.

\paragraph*{High-level proof overview} Our technique extends and builds upon
the meta-theorem of \cite{Gajarsky15}, which handles the more restricted case
of graphs of bounded tree-depth. We recall that the heart of this meta-theorem
is the basic observation that FO logic has bounded counting power: if our graph
contains $q+1$ identical parts (for some appropriate definition of
``identical''), then deleting one cannot affect the validity of any FO formula
with $q$ quantifiers. The approach of \cite{Gajarsky15} is to partition the
vertices of the graph depending on their height in the tree-depth
decomposition, then identify (and delete) identical vertices in the bottom
level. This bounds the degree of vertices one level up, which allows us to
partition them into a bounded number of types, delete components of the same
type if we have too many, hence bound the degree of vertices one level up, and
so on until the size of the whole graph is bounded.

Our approach borrows much of this general strategy: we will appropriately rank
the vertices of the graph and then move from lower to higher ranks, at each
step bounding the maximum degree of any vertex of the current rank. Besides the
fact that ranking vertices into levels is less obvious when given a path
decomposition, rather than a tree of fixed height, the main difficulty we
encounter is that no matter where we start, we cannot in general easily find
identical parts where something can be safely deleted.  Intuitively, this is
demonstrated by the contrast between the simplest bounded tree-depth graph (a
star, where leaves are twins, hence one can easily delete one if we have at
least $q+1$) and the simplest bounded pathwidth graph (a path, which contains
no twins). In order to handle this more general case, we need to combine the
previous approach with arguments that rely on the locality of FO logic.

To understand informally what we mean by this, recall the classical argument
which proves that \textsc{Reachability} is not expressible in FO logic. One way
this is proved is to show that a graph $G_1$ which is a long path (of say,
$4^q$ vertices) and a graph $G_2$ which is a union of a path and a cycle (of
say, $2\cdot 4^{q-1}$ vertices each) are indistinguishable for FO formulas with
$q$ quantifiers. Our strategy is to flip this argument: if we are asked to
model check a formula on a long path, we might as well model check the same
formula on a simpler (less connected) graph which contains a shorter path and a
cycle.  Of course, our input graphs will be more complicated than long paths;
we will, however, be dealing with long path-like structures, as our graph has
small pathwidth. Our strategy is to perform a surgical rewiring operation on
the path decomposition, producing the union of a shorter decomposition and a
ring-like structure, while still satisfying the same formulas (the reader may
skip ahead to Figure~\ref{fig:rewiring} to get a feeling for this operation). In
other words, the main technical ingredient of our algorithm is inspired by (and
exploits) a classical inexpressibility result for FO logic.  The abstract idea
is (in a rough sense) to apply this argument repeatedly, so that if we started
with a long path decomposition, we end up with a short path decomposition and
many ``disconnected rings''. Eventually, we will be able to produce some such
rings which are identical, delete them, and simplify the graph.

There are, of course, now various technical difficulties we need to overcome in
order to turn this intuition into a precise argument. First, when we cut at two
points in the path decomposition to extract the part that will form the
``ring'', we need to make sure that at an appropriate radius around the cut
points the decompositions are isomorphic.  It is not hard to calculate the
appropriate radius we need in the graph (it is known that $q$-quantifier FO
formulas depend on balls of radius roughly $2^q$), but a priori two vertices
which are close in the graph could be far in the path decomposition. To handle
this, we take care when we rank the vertices, so that vertices of lower rank
are guaranteed to only appear in a bounded number of bags, hence distances in
the path decomposition approximate distances in the graph. Second, we need to
calculate how long our decomposition needs to be before we can guarantee that
we will be able to find some appropriate cut points. Here we use some counting
arguments and pigeonhole principle to show that a path decomposition with
length double-exponential in the desired radius is sufficient. Finally, once we
find sufficiently many points to rewire and produce sufficiently many
``rings'', we need to prove that this did not affect the validity of the
formula. Then, we are free to delete one, using the same argument as
\cite{Gajarsky15} and obtain a smaller equivalent graph. In the end, once we
can no longer repeat this process, we obtain a bounded-degree graph, where it
is known that FO model checking has an elementary dependence on the formula.

Overall, even though the algorithm we present seems somewhat complicated, the
basic ingredients are simple and well-known: the fact that deleting one of many
identical parts does not affect the validity of the formula (which is also used
in \cite{Gajarsky15}); the fact that FO formulas are not affected if we edit
the graph in a way that preserves balls of a small radius around each vertex;
and simple counting arguments and the pigeonhole principle. 

\paragraph*{Paper organization} We conclude this section below with a short
overview of other related work on algorithmic meta-theorems and continue in
Section~\ref{sec:defs} with definitions and notation. The rest of the paper is
organized as follows:

\begin{enumerate}

\item In Section~\ref{sec:basic} we present two lemmas, which are standard facts on FO
logic, with minor adjustments to our setting. In particular, in Section~\ref{sec:id}
we present the lemma that states that if we have $q+1$ identical parts, it is
safe to delete one; and in Section~\ref{sec:sameN} we present the lemma that states
that if two graphs agree on the local extended neighborhoods around each vertex
(for some appropriate radius), then they satisfy the same formulas (that is, FO
logic is local). Since these facts are standard, the reader may wish to skip
the proofs of Section~\ref{sec:basic}, which are given for the sake of completeness,
during a first reading.

\item Then, in Section~\ref{sec:operations} we present the specific tools we will use
to simplify our graph. In Section~\ref{sec:normal} we explain how we rank the vertices
of a path decomposition so that each vertex has few neighbors of higher rank
(but possibly many neighbors of lower rank). This allows us to process the
ranks from lower to higher, simplifying the graph step by step.  Then, in
Section~\ref{sec:iso} we use some counting arguments to calculate the length of a path
decomposition that guarantees the existence of long isomorphic blocks, on which
we will apply the rewiring operation. We also show how distances in the graph
can be approximated by distances in the path decomposition if we have bounded
the number of occurrences of each vertex in the decomposition. Finally, in
Section~\ref{sec:rewiring} we formally define the rewiring operation and show that if
the points where we apply it are in the middle of sufficiently long isomorphic
blocks of the decomposition, this operation is safe. We also show that the
``rings'' it produces can be considered identical, in a sense that will allow
us to invoke the lemma of Section~\ref{sec:id} and delete one.

\item We put everything together in Section~\ref{sec:final}, where we explain how the
lemmas we have presented form parts of an algorithm that ranks the vertices of
a graph supplied with a path decomposition and then processes ranks one by one,
decreasing the number of occurences of each vertex in the decomposition without
affecting the validity of any formula (with $q$ quantifiers). In the end, the
processed graph has bounded degree and we invoke known results to decide the
formula.

\end{enumerate}

\paragraph*{Other related work} Algorithmic meta-theorems are a very
well-studied topic in parameterized complexity and much work
has been devoted in improving and extending Courcelle's theorem. Among such
results, we mention the generalization of this theorem to MSO for clique-width,
which covers dense graphs \cite{CourcelleMR00}. For FO logic, fixed-parameter
tractability can be extended to much wider classes of graphs, with the recently
introduced notion of twin-width nicely capturing many results in the area
\cite{BonnetKTW22,DvorakKT13,Frick04,FrickG01}. Of course, since all these
classes include the class of all trees, the non-elementary dependence on the
formula implied by the lower bound of \cite{FrickG04} still applies.
Meta-theorems have also been given for logics other than FO and MSO, with the
goal of either targeting a wider class of problems
\cite{GanianO13,KnopKMT19,KnopMT19,Szeider11}, or achieving better complexity
\cite{Pilipczuk11}.  Kernelization
\cite{BodlaenderFLPST16,EibenGS18,GanianSS16} and approximation
\cite{DawarGKS06} are also topics where meta-theorems have been studied.

The meta-theorems which are more relevant to the current work are those which
explicitly try to improve upon the parameter dependence given by Courcelle, by
considering more restricted parameters. We mention here the meta-theorems for
vertex cover, max-leaf,  and neighborhood diversity \cite{Lampis12}, twin-cover
\cite{Ganian15}, shrub-depth \cite{GanianHNOM19}, and vertex integrity
\cite{LampisM24}. As mentioned, one common aspect of these meta-theorems is
that they handle both FO and MSO logic, without a huge difference in complexity
(at most one extra level of exponentiation in the parameter dependence), which
makes the behavior of FO logic on pathwidth somewhat unusual. The only
exception, is the meta-theorem on graphs of bounded max-leaf number of
\cite{Lampis12} which does not generalize to MSO logic. It was later shown that
this is with good reason, as MSO logic has a non-elementary dependence even for
unlabeled paths \cite{Lampis14}, which have the smallest possible max-leaf
number. This is therefore the only previous result in the literature which
mirrors the situation for pathwidth.

A classical result, incomparable to the parameters mentioned above, is the fact
that FO model checking is FPT (with an elementary, triple-exponential
dependence on the formula) on graphs of bounded degree \cite{Seese96}. We will
use this fact as the last step of our algorithm.

The complexity of model checking FO and MSO formulas on structures other than
graphs, such as posets \cite{GajarskyHLOORS15} and strings has also been
investigated. As mentioned, the case of strings is of particular interest to
us, because the standard structure that represents a string over a fixed
alphabet (a universe that contains the letters of the string, unary predicates
that indicate for each letter which character of the alphabet it corresponds
to, and a total ordering relation $\prec$ which indicates the ordering of the
letters in the string) allows us to easily translate MSO properties of strings
into MSO properties of an appropriate caterpillar.  Indeed, to embed a string
into a caterpillar, we can start with a path with endpoints $s,t$, and use one
vertex of the path to represent each letter in the string. We can attach an
appropriate (constant) number of leaves on each vertex to signify which
character it represents. The precedence relation $x\prec y$ of the string now
becomes the relation ``every connected set that contains $s$ and $y$, also
contains $x$'', which is MSO-expressible. Thanks to this simple transformation,
the lower bound result of \cite{FrickG04} on model checking MSO (and even FO)
logic on strings immediately carries over to graphs of pathwidth $1$. Note,
however, that the existence of the ordering relation is crucial, as FO model
checking on other models of strings (e.g. with a successor relation) has
elementary dependence on the formula, as such structures have bounded degree
\cite{FrickG04}. Hence, it seems that if we focus on FO (rather than MSO)
logic, the similarity between model checking on bounded pathwidth graphs and
strings becomes much weaker: FO model checking is easier on graphs of bounded
pathwidth than on strings with an ordering relation, but harder than on strings
with only a successor relation (as the lower bound of \cite{Lampis14} for
tree-depth applies to pathwidth, and rules out an algorithm with ``only''
triple-exponential dependence).

\paragraph*{Update on related work} Subsequently to the conference version
of this paper, recent work has appeared (\cite{GajarskyP0ST24}) which
significantly generalizes our results, extending them to the realm of
nowhere-dense graphs. More precisely, \cite{GajarskyP0ST24} gives a precise
characterization of the subgraph-closed classes on which FO model checking can
be performed with an elementary dependence on the formula and their result
includes bounded pathwidth graphs as a special case. Their work builds upon the
algorithm presented here, but requires many other ideas and is significantly
more involved and technical.

\section{Definitions and Preliminaries}\label{sec:defs}

We use standard graph-theoretic notation and assume the reader is familiar with
the basics of parameterized complexity (see e.g. \cite{CyganFKLMPPS15}). For a
graph $G=(V,E)$, and $S\subseteq V$, we use $G[S]$ to denote the subgraph of
$G$ induced by $S$. When $r$ is a positive integer, we use $[r]$ to denote the
set $\{1,\ldots,r\}$, while for two integers $s,t$, we use $[s,t]$ to denote
the set $\{ i\in\mathbb{Z}\ |\ s\le i\le t\}$. Note that if $t<s$ then
$[s,t]=\emptyset$. We define $\tow(i,n)$ as follows: $\tow(0,n)=n$ and
$\tow(i+1,n)=2^{\tow(i,n)}$. A function $f:\mathbb{N}\to\mathbb{N}$ is
elementary if there exists a fixed $i$ such that for all $n$ we have $f(n)\le
\tow(i,n)$.

We recall the standard notion of path decomposition: a path decomposition of a
graph $G=(V,E)$ is an ordered sequence of bags $B_1,B_2,\ldots,B_{\ell}$, where
each $B_i$ is a subset of $V$ that satisfies the following: (i)
$\bigcup_{j\in[\ell]}B_j=V$ and for all $uv\in E$ there exists $i\in[\ell]$
such that $\{u,v\}\subseteq B_i$ (ii) for all $i_1<i_2<i_3$, with
$i_1,i_2,i_3\in [r]$ we have $B_{i_1}\cap B_{i_3}\subseteq B_{i_2}$. The width
of a path decomposition is the number of vertices in the largest bag (minus
one). The pathwidth of a graph $G$ is the smallest width of any path
decomposition of $G$.

\paragraph*{First Order Logic} We use a standard form of First Order (FO)
logic on graphs, where quantified variables are allowed to range over vertices.
To simplify the presentation of some results, we will allow our formulas to
also refer to vertex constants, corresponding to some specific vertices of the
graph. More formally, the structures on which we will perform model checking
are $k$-terminal graphs as defined below.

\begin{defi}\label{def:kterm} For a positive integer $k$, a $k$-terminal
graph $G=(V,E)$ is a graph supplied with a function $\mathcal{T}: [k]\to V$,
called the terminal labeling function.  For $i\in [k]$, we say that
$\mathcal{T}(i)$ is the $i$-th terminal of $G$.  The set of terminals is the
set $T$ of images of $\mathcal{T}$ in $V$.  Vertices of $V\setminus T$ are
called non-terminals.  \end{defi} 

Intuitively, terminals will play two roles: on the one hand, we define FO logic
on graphs (below) in a way that allows formulas to refer to the terminal
vertices; on the other, in some parts of our algorithm we will use a set of
terminals that form a separator of the graph and hence allow us to break down
the graph into smaller components. Note, however, that Definition~\ref{def:kterm} does
not require the $k$ terminals to be a separator, or have any other particular
property.

A formula of FO logic is made up of the following vocabulary: (i) vertex
variables, denoted $x_1,x_2,\ldots$ (ii) vertex constants denoted
$\ell_1,\ell_2,\ldots$ (iii) existential quantification $\exists$ (iv) the
Boolean operations $\neg, \lor$ (v) the binary predicates $\sim$ (for
adjacency) and $=$ (for equality). More formally, a First Order formula is a
formula produced by the following grammar, where $x$ represents a vertex
variable and $y$ represents a vertex variable or constant:

\[ \phi \to \exists x. \phi\ |\ \neg \phi\ |\ (\phi \lor \phi)\ |\ y\sim y\ |\
y=y \]

A FO formula $\phi$ is called a sentence if every vertex variable $x$ appearing
in $\phi$ is quantified, that is, $x$ appears within the scope of $\exists x$.
A variable that is not quantified is called a free variable. For a formula
$\phi$ that contains a free variable $x$, we will write $\phi[x/\ell_i]$ to
denote the formula obtained by replacing every occurrence of $x$ in $\phi$ by
the constant $\ell_i$.

The main problem we are concerned with is model checking: given a $k$-terminal
graph $G$ and a sentence $\phi$, decide if $G$ satisfies $\phi$. We define the
semantics of what this means inductively in a standard way, as follows.  We say
that a $k$-terminal graph $G=(V,E)$ with labeling function $\mathcal{T}$ models
(or satisfies) a sentence $\phi$ (using only constants $\ell_i$ with
$i\in[k]$), and write $G,\mathcal{T}\models\phi$ (or simply $G\models\phi$ if
$\mathcal{T}$ is clear from the context) if and only if we have one of the
following:

\begin{enumerate}

\item $\phi:=\ell_i = \ell_j$, where $i,j\in [k]$ and $\mathcal{T}(i)$ is the
same vertex as $\mathcal{T}(j)$.

\item $\phi:=\ell_i \sim \ell_j$, where $i,j\in[k]$ and
$\mathcal{T}(i)\mathcal{T}(j)\in E$.

\item $\phi:=\neg \psi$ and it is not the case that $G,\mathcal{T}\models
\psi$.

\item $\phi:=(\psi_1\lor \psi_2)$ and at least one of
$G,\mathcal{T}\models\psi_1$, $G,\mathcal{T}\models\psi_2$ holds.

\item $\phi:=(\exists x. \psi)$ and there exists $v\in V$ such that
$G,\mathcal{T}'\models \psi[x/\ell_{(k+1)}]$, where $\mathcal{T}'$ is the
labeling function that sets $\mathcal{T}'(k+1)=v$ and
$\mathcal{T}'(i)=\mathcal{T}(i)$ for $i\in [k]$. 

\end{enumerate}

Note that we have not included in our definition universal quantification or
other Boolean connectives such as $\land$. However, this is without loss of
generality as $\forall x.\phi$ can be thought of as shorthand for $\neg \exists
x.\neg \phi$ and all missing Boolean connectives can be simulated using $\neg$
and $\lor$.

Let us also define a kind of isomorphism between labeled graphs that is
guaranteed to leave terminal vertices untouched.

\begin{defi} A terminal-respecting isomorphism between two $k$-terminal
graphs $G_1=(V_1,E_1)$ and $G_2=(V_2,E_2)$ with terminal labeling functions
$\mathcal{T}_1,\mathcal{T}_2$ is a bijective function $f:V_1\to V_2$ such that
(i) for all $u,u'\in V_1$ we have $uu'\in E_1$ if and only if $f(u)f(u')\in
E_2$ (ii) for each $i\in[k]$, $f(\mathcal{T}_1(i)) = \mathcal{T}_2(i)$.
\end{defi} 

We recall the following basic fact about FO logic which states that isomorphic
structures satisfy the same sentences (this also appears as Lemma 2.9 in
\cite{LampisM24} and a proof can easily be obtained by structural induction).

\begin{lem}\label{lem:iso1} If $G_1,G_2$ are two $k$-terminal graphs such
there exists a terminal-respecting isomorphism from $G_1$ to $G_2$, then, for
all $FO$ sentences $\phi$ we have $G_1\models \phi$ if and only if $G_2\models
\phi$.  \end{lem}

\section{Two Basic Lemmas}\label{sec:basic}

The purpose of this section is to establish two basic ingredients that will
allow us to simplify the input graph without affecting whether it satisfies any
FO formula with at most a given number $q$ of quantified variables. The first
lemma (Lemma~\ref{lem:isomorphic}) is rather simple and states that if a graph
contains many ``identical'' components, we can safely remove one. Despite its
simplicity, this idea has been sufficient to obtain many of the best currently
known meta-theorems with non-elementary dependence in the formula, such as the
meta-theorem of \cite{Gajarsky15} for graphs of bounded tree-depth. 

The second lemma (Lemma~\ref{lem:sameN}) is a variation of a standard fact regarding
the locality of FO logic known as Hanf's locality lemma. It states that if we
have two graphs which look locally the same, in the sense that for each vertex
of one graph there exists a vertex of the other whose $r$-neighborhood is the
same, for some appropriately chosen $r$, then actually the two graphs are
indistinguishable by FO formulas with $q$ quantifiers (even though they are not
necessarily isomorphic).  As we explained, we intend to use this to allow us to
take parts of the graph that resemble ``long'', low-pathwidth components and
cut them up into smaller, disconnected components.  The strategy is to
eventually produce a large enough number of such components that we can apply
Lemma~\ref{lem:isomorphic} and simplify the graph.

It is useful to stress here that the results of this section are not new --
indeed all the lemmas we present here are well-known. We prefer to give full
proofs for the sake of a self-contained presentation consistent with our
conventions. Given this, a knowledgeable reader may feel free to skip the
proofs of this section.

\subsection{Identical Parts}\label{sec:id}

We would now like to show that if the given graph contains many (say, at least
$q+1$) ``identical'' parts, then it is safe to delete one without affecting
whether the graph satisfies any FO formula with at most $q$ quantifiers.  We
first define what we mean that two sets of vertices are identical in a
$k$-terminal graph and then prove that if we can find $q+1$ such sets in a
graph, we can safely delete one without affecting whether any FO formula with
at most $q$ quantifiers is satisfied.

\begin{defi}\label{def:id} Let $G=(V,E)$ be a $k$-terminal graph with
labeling function $\mathcal{T}$ and terminal set $T$. We say that two disjoint
sets of vertices $C_1,C_2$ are \emph{identical} if there exists a
terminal-respecting isomorphism from $G$ to $G$ that maps all vertices of $C_1$
to $C_2$ and all vertices of $C_2$ to $C_1$, and maps every vertex of
$V\setminus (C_1\cup C_2)$ to itself.  \end{defi}

Before we proceed, let us make two easy observations. First, if $C_1,C_2$ are
identical, it must be the case that $(C_1\cup C_2)\cap T=\emptyset$, because
$C_1,C_2$ are disjoint and terminal-respecting isomorphisms must map vertices
of $T$ to themselves. Second, the relation of being identical is an equivalence
relation on a collection of pairwise disjoint sets of vertices, that is, if
$C_1,C_2,C_3$ are disjoint, $C_1$ is identical to $C_2$, and $C_2$ is identical
to $C_3$, then $C_1$ is identical to $C_3$.

\begin{lem}\label{lem:isomorphic} Fix a positive integer $q$. Let $G=(V,E)$
be a $k$-terminal graph with labeling function $\mathcal{T}$ and terminal set
$T$ and suppose that $C_1,C_2,\ldots,C_{q+1}$ are $q+1$ disjoint sets of
vertices of $G$ which are pairwise identical. Then, for all FO sentences with
at most $q$ quantifiers we have that $G,\mathcal{T}\models \phi$ if and only if
$G[V\setminus C_1],\mathcal{T}\models \phi$.  \end{lem}

\begin{proof} 

We prove the lemma by structural induction on the formula $\phi$.

\begin{enumerate}

\item If $\phi := (\ell_i=\ell_j)$ or $\phi := (\ell_i\sim \ell_j)$, then
whether the formula $\phi$ is satisfied only depends on the graph induced by
the terminal vertices $T$ and the labeling function $\mathcal{T}$. Since both
are the same in $G$ and $G[V\setminus C_1]$, the base case holds.

\item If $\phi := (\neg \psi_1)$ or $\phi = (\psi_1\lor\psi_2)$, then we can
assume that by inductive hypothesis $\psi_1,\psi_2$ do not distinguish between
$G, G[V\setminus C_1]$. Hence, if $G\models \psi_1$ then $G[V\setminus
C_1]\models \psi_1$, and therefore both $G, G[V\setminus C_1]$ satisfy
$\psi_1\lor \psi_2$ (and similarly for the case of negation).

\item Finally, if $\phi := (\exists x. \psi)$ we first prove that if
$G[V\setminus C_1],\mathcal{T}\models \phi$ then $G,\mathcal{T}\models \phi$.
In this case, there must exist $v\in V\setminus C_1$ such that $G[V\setminus
C_1],\mathcal{T}' \models \psi[x/\ell_{(k+1)}]$, where $\mathcal{T}'$ is the
labeling function that agrees with $\mathcal{T}$ on $i\in [k]$ and sets
$\mathcal{T}'(k+1)=v$. We now observe that we can apply the inductive
hypothesis on $\psi, G, G[V\setminus C_1]$ for the new labeling function
$\mathcal{T}'$, because even if $v\in \bigcup_{i=2\ldots q+1} C_i$, there still
exist $q$ identical components among $C_1,\ldots, C_{q+1}$ in $G, G[V\setminus
C_1]$ and $\psi$ has $q-1$ quantifiers. By the inductive hypothesis we have
$G,\mathcal{T}'\models \psi[x/\ell_{(k+1)}]$, which implies that
$G,\mathcal{T}\models \phi$.

For the converse direction, suppose $G,\mathcal{T}\models \phi$ and we want to
prove that in this case $G[V\setminus C_1],\mathcal{T}\models\phi$. Again, by
definition there exists $v\in V$ such that $G,\mathcal{T}'\models
\psi[x/\ell_{(k+1)}]$, where $\mathcal{T}'$ is the labeling function that sets
$\mathcal{T}'(k+1)=v$ and agrees with $\mathcal{T}$ elsewhere. Now, if
$v\not\in C_1$, then we use the inductive hypothesis, as in the previous
paragraph, to argue that $G[V\setminus C_1],\mathcal{T}'\models
\psi[x/\ell_{(k+1)}]$ and hence $G[V\setminus C_1]\models \phi$. Suppose then
that $v\in C_1$. We recall that there exists a terminal-respecting isomorphism
$f:V\to V$ that maps $C_1$ to $C_2$ and all other vertices to themselves, since
$C_1,C_2$ are identical. Let $u=f(v)$. We claim that for the labeling function
$\mathcal{T}''$ which sets $\mathcal{T}''(k+1)=u$ and
$\mathcal{T}''(i)=\mathcal{T}(i)$ for $i\in [k]$, we have
$G,\mathcal{T}''\models \psi[x/\ell_{(k+1)}]$. Indeed, this is a consequence of
Lemma~\ref{lem:iso1} and the fact that $f$ is a terminal-respecting isomorphism from
$G,\mathcal{T}'$ (which satisfies the same formula) to $G,\mathcal{T}''$.
Hence, we have reduced this to the case where $v\not\in C_1$ and we are done. \qedhere

\end{enumerate} \end{proof}

\subsection{Similar Neighborhoods}\label{sec:sameN}

We now move on to present a lemma that will allow us to claim that two graphs
are indistinguishable for FO formulas with $q$ quantifiers if they are locally
the same. This is a standard argument in FO logic, going back to Gaifman,
though we need to adjust the proof to our purposes to handle terminal vertices
appropriately. Our version will use essentially the same proof as that given in
Theorem 4.12 of \cite{Libkin04}, but with some minor technical differences. In
particular, we will on the one hand be stricter on the isomorphisms we allow
before we consider that the neighborhoods of two vertices are the same (because
we only allow terminal-respecting isomorphisms), but on the other, we will only
consider the extended neighborhood around a vertex by considering paths that go
through non-terminals. This is important, because it allows us to work around
the case where, for example, a terminal vertex is connected to everything and
hence the diameter of the graph is $2$.  In such a case, the extended
neighborhood of a non-terminal vertex will not trivially contain the whole
graph, because we exclude paths that go through the supposed universal
terminal.

According to this discussion, we define the notion of a ball of radius $r$
around a vertex $v$, denoted $B_r(v)$, in a way that only takes into account
paths whose internal vertices are non-terminals, as follows.

\begin{defi} Let $G=(V,E)$ be a $k$-terminal graph with terminal labeling
function $\mathcal{T}$ and terminal set $T$, $r$ be a positive integer, and
$v\in V$.  We define $B^G_r(v)$ (and simply write $B_r(v)$ if $G,\mathcal{T}$
are clear from the context) to be the $k$-terminal subgraph of $G$ that has
labeling function $\mathcal{T}$ and is induced by $T\cup V'$, where $V'$ is the
set of all vertices reachable by $v$ via a path of length at most $r$ whose
internal vertices are all in $V\setminus T$.  \end{defi}

\begin{defi}\label{def:similar} Let $G_1=(V_1,E_1), G_2=(V_2,E_2)$ be two
$k$-terminal graphs, with terminal labeling functions
$\mathcal{T}_1,\mathcal{T}_2$ and terminal sets $T_1,T_2$.  For a non-negative
integer $r$, we will say that $v_1\in V_1$ is $r$-\emph{similar} to $v_2\in
V_2$, if there exists a terminal-respecting isomorphism from $B^{G_1}_r(v_1)$
to $B^{G_2}_r(v_2)$ that maps $v_1$ to $v_2$.  \end{defi}

Note that in the above definition, $G_1$ and $G_2$ may be the same graph. It is
not hard to see that $r$-similarity is an equivalence relation on the vertices
of $V_1\cup V_2$. Definition~\ref{def:similar} allows us to set $r=0$, in which case we
are testing if the graphs induced by $T\cup\{v_1\}$ and $T\cup\{v_2\}$ are
isomorphic. Let us also make the following easy observation that decreasing $r$
cannot make two similar vertices dissimilar.

\begin{obs}\label{obs:easy} Let $G_1,G_2$ be two graphs as in
Definition~\ref{def:similar} and $v_1\in V(G_1), v_2\in V(G_2)$ be two vertices which are
$r$-similar. Then, for all non-negative integers $r'\le r$, $v_1$ is
$r'$-similar to $v_2$.  \end{obs}

\begin{proof} We prove the observation for $r>0$ and $r'=r-1$, as this is
sufficient. Since $v_1,v_2$ are $r$-similar, there exists a terminal-respecting
isomorphism $f:B_r^{G_1}(v_1) \to B_r^{G_2}(v_2)$. We use the same isomorphism,
but restricted to $B_{r'}^{G_1}(v_1)$. The image of $f$ is now
$B_{r'}^{G_2}(v_2)$, because (induced subgraph) isomorphisms preserve distances
and it is not hard to see that all other requirements are also satisfied.
\end{proof}

The main lemma of this section is then the following.

\begin{lem}\label{lem:sameN} Let $q,k$ be positive integers and set
$r=2^q-1$.  Let $G_1,G_2$ be two $k$-terminal graphs that contain some
non-terminal vertices, with labeling functions $\mathcal{T}_1, \mathcal{T}_2$
and terminal sets $T_1,T_2$.  Suppose that there exists a bijective mapping
$f:V(G_1)\to V(G_2)$ such that (i) for all $i\in[k]$ we have
$f(\mathcal{T}_1(i))=\mathcal{T}_2(i)$ (ii) for all non-terminal vertices $v\in
V(G_1)\setminus T_1$ we have that $v$ is $r$-similar to $f(v)\in
V(G_2)\setminus T_2$.  Then, for all FO formulas $\phi$ with at most $q$
quantifiers we have $G_1\models \phi$ if and only if $G_2\models \phi$.
\end{lem}

\begin{proof}

We prove the statement by induction on $q$. For $q=0$ the graph induced by the
terminals of $G_1$ is isomorphic to the graph induced by the terminals of
$G_2$.  To see this, take any non-terminal vertex $v_1\in V(G_1)$ and
$v_2=f(v_1)\in V(G_2)$ and note that, since by assumption $v_1,v_2$ are
$r$-similar (and $r=0$), we have that $G_1[T_1\cup\{v_1\}]$ and
$G_2[T_2\cup\{v_2\}]$ are isomorphic.  Hence, $G_1,G_2$ satisfy the same
quantifier-free FO formulas, because non-terminal vertices are irrelevant for
such formulas.

Suppose now that $q>0$ and the statement is true for formulas with at most
$q-1$ quantifiers.  Consider a formula $\phi$ with $q$ quantifiers and suppose
that $\phi=\exists x.  \psi(x)$. We claim that this is without loss of
generality: if $\phi$ has the form $\forall x. \psi(x)$ we could apply our
argument to $\neg \phi = \exists x. (\neg \psi(x))$, while if $\phi$ is a
Boolean combination of other formulas we can consider each subformula
separately.

Suppose that $G_1,\mathcal{T}_1\models \phi$. We will show that this implies
that $G_2,\mathcal{T}_2\models \phi$. Since we can inverse the roles of
$G_1,G_2$ by taking the inverse mapping $f^{-1}$, by symmetry this will also
establish the converse implication and hence the lemma.

Since $G_1,\mathcal{T}_1\models \exists x. \psi(x)$, there must exist a vertex
$v_1\in V_1$ such that $G_1,\mathcal{T}_1'\models \psi[x/\ell_{(k+1)}]$, where
$\mathcal{T}_1'(i)=\left\{ \begin{array}{ll} \mathcal{T}_1(i) & \textrm{if }
i\le k \\ v_1& \textrm{otherwise}\end{array}\right.  $. Let $v_2\in V_2$ be the
vertex for which $f(v_1)=v_2$.  Define $\mathcal{T}_2'(i)=\left\{
\begin{array}{ll} \mathcal{T}_2(i) & \textrm{if } i\le k \\ v_2&
\textrm{otherwise}\end{array}\right.  $. We claim that we can apply the
inductive hypothesis to  show that $G_2,\mathcal{T}_2'\models
\psi[x/\ell_{(k+1)}]$ and hence that $G_2,\mathcal{T}_2\models \phi$, as
desired.

To prove this claim, we first distinguish the easy case where $v_1\in T_1$, and
hence $v_2\in T_2$. In this case, for all non-terminal vertices $v$ the ball
$B_r(v)$ remains unchanged in both graphs, as the set of terminals is
unchanged. Hence, if a non-terminal $v\in V(G_1)$ was $r$-similar to $f(v)$,
$v$ is still $r'$-similar to $v$, for $r'=2^{q-1}-1<r$ under the new labelings
$\mathcal{T}_1', \mathcal{T}_2'$ (using Observation~\ref{obs:easy}). Hence, since $\psi$
has $q-1$ quantifiers, we can apply the inductive hypothesis using the same
bijective mapping $f$.

For the more interesting case where $v_1\not\in T_1$, we note that $v_2\not\in
T_2$ is $r$-similar to $v_1$, for $r=2^q-1$, that is, $B_r^{G_1}(v_1)$ has a
terminal-respecting isomorphism to $B_r^{G_2}(v_2)$, where the balls are with
respect to $\mathcal{T}_1, \mathcal{T}_2$ respectively.

To show that the inductive hypothesis applies to $G_1,G_2$, which are now
$(k+1)$-terminal graphs for the labeling functions
$\mathcal{T}_1',\mathcal{T}_2'$, we need to produce a new bijective mapping
$f':V(G_1)\to V(G_2)$. Setting $f'(v)=f(v)$ for $v\in T_1\cup\{v_1\}$ satisfies
the first necessary property of the mapping, so what remains is to take care of
non-terminals. We therefore need the property that for all non-terminals $v\in
V(G_1)$, $v$ is $r'$-similar to $f'(v)$ (with respect to the labeling functions
$\mathcal{T}_1',\mathcal{T}_2'$), where $r'=2^{q-1}-1$, taking into account the
new terminal which has been added to the labeling functions.  In order to do
so, we construct a bipartite graph $H$ with vertex set $V(G_1)\cup V(G_2)$.
Recall that we were initially given a mapping $f:V(G_1)\to V(G_2)$; for each
$v\in V(G_1)$ we add the edge $vf(v)$ in $H$.  Call the edges we added so far
blue edges.  Furthermore, recall that $v_2=f(v_1)$ is $r$-similar to $v_1$,
therefore by definition there exists a terminal-respecting isomorphism, call it
$f^*:B_r^{G_1}(v_1)\to B_r^{G_2}(v_2)$ (where here the balls are with respect
to $\mathcal{T}_1, \mathcal{T}_2$).  Note that $f^*$ is terminal-respecting
also if we take into account the new terminal, as $f^*(v_1)=v_2$. For a vertex
$v\in B_r^{G_1}(v_1)$ we will say that $v$ is close to $v_1$ if there exists a
path of length at most $r'+1$ connecting $v$ to $v_1$ in $G_1$ without using
any internal terminal vertices.  For each $v\in B_r^{G_1}(v_1)$ that is close
to $v_1$ we add in $H$ the edge $vf^*(v)$.  Call such edges red edges.

Observe now that in $H$ every vertex is incident to exactly one blue edge and
at most one red edge. Hence, every connected component of $H$ is either an even
cycle with the same number of red and blue edges, or an alternating blue-red
path, where the first and last edges are blue. We define the mapping $f'$ as
follows: for each $v\in V(G_1)$ which is incident to a red edge, we set $f'(v)$
to be the other endpoint of that edge; for each remaining $v\in V(G_1)$, we
observe that $v$ must be an endpoint of an alternating blue-red path in $H$,
and we set $f'(v)$ to be the other endpoint of that path. It is not hard to see
that this mapping is indeed one-to-one.

To complete the proof we need to show that the mapping $f'$ matches vertices
which are $r'$-similar in $G_1, G_2$ with respect to the new labelings
$\mathcal{T}_1', \mathcal{T}_2'$.  Consider first a $v\in V(G_1)$ which is
incident to a red edge of $H$. Hence, $v$ is close (that is, at distance at
most $r'+1$) to $v_1$ in $B_r^{G_1}(v_1)$ and $f'(v)=f^*(v)$. We now observe
that $B_{r'}^{G_1}(v)$ is fully contained in $B_r^{G_1}(v_1)$, because any
vertex that is at distance at most $r'$ from $v$ is at distance at most
$r'+(r'+1)=r$ from $v_1$. Because $f^*$ is an isomorphism from $B_r^{G_1}(v_1)$
to $B_r^{G_2}(v_2)$, we have that $B_{r'}^{G_1}(v)$ is isomorphic to
$B_{r'}^{G_2}(f^*(v))$. Hence, $v$ and $f'(v)$ are $r'$-similar in the new
labeled graphs, as desired.

Finally, consider a vertex $v\in V(G_1)$ which is incident only to a blue edge
of $H$ and has been matched to the other endpoint of the alternating blue-red
path starting at $v$ (note that this path could consist of a single blue edge).
Suppose the other endpoint of this path is $v'\in V(G_2)$. We want to show that
$v,v'$ are $r'$-similar, taking into account the new terminal. We first observe
that any two vertices which are connected by a blue edge in $H$ are $r$-similar
in $G_1,G_2$ with the old labeling functions $\mathcal{T}_1,\mathcal{T}_2$.
Such vertices are also $r'$-similar with respect to
$\mathcal{T}_1,\mathcal{T}_2$, as $r'\le r$, by Observation~\ref{obs:easy}.  Furthermore,
vertices connected by a red edge are also $r'$-similar in $G_1,G_2$, because,
as we argued in the previous paragraph, a ball of radius $r'$ around such a
vertex is fully contained in $B_r^{G_1}(v_1)$ or $B_r^{G_2}(v_2)$. Recall that
$r'$-similarity is an equivalence relation.  As a result, $v,v'$ are
$r'$-similar in $G_1,G_2$. To conclude that they are still similar after adding
the extra terminal, we observe that both $v,v'$ are at distance at least $r'+2$
from $v_1,v_2$ respectively (otherwise they would have been incident to a red
edge), hence in both $B_{r'}^{G_1}(v)$ and $B_{r'}^{G_2}(v')$, the $(k+1)$-th
terminal is disconnected from all non-terminal vertices.  \end{proof}

\section{Simplification Operations on Path
Decompositions}\label{sec:operations}

In this section we present the main technical ingredients of our algorithm. In
Section~\ref{sec:normal} we show how we can rank the vertices to bound the number of
higher-rank neighbors of any vertex; in Section~\ref{sec:iso} we use the pigeonhole
principle to show that for sufficiently long path decompositions we can always
find long isomorphic blocks; and in Section~\ref{sec:rewiring} we describe the
rewiring operation we will use in these blocks and show that it does not affect
the validity of any formula and that it produces identical parts, in the sense
of Lemma~\ref{lem:isomorphic}.

\subsection{Normalized Path Decompositions}\label{sec:normal}

\begin{defi}\label{def:rank} A \emph{ranked} path decomposition of a
graph $G=(V,E)$ is a path decomposition together with a ranking function
$\rho:V\to \mathbb{N}$ that has the property that no bag $B_i$ of the
decomposition contains two vertices $u,v\in B_i$ for which
$\rho(u)=\rho(v)$.\end{defi}

Given a path decomposition, it is in general easy to produce a ranking function
as defined in Definition~\ref{def:rank}. We will, however, require a ranking function
with some useful properties, as stated in the following lemma.

\begin{lem}\label{lem:normal} Given a graph $G=(V,E)$ and a path
decomposition of $G$ where each bag contains at most $p$ vertices, it is
possible in polynomial time to convert it into a ranked path decomposition with
a ranking function $\rho:V\to [8p]$ and the property that for each $i<j$ with
$i,j\in [8p]$ we have that for every vertex $v$ with $\rho(v)=i$, there exist at
most two vertices $u_1,u_2$ with $\rho(u_1)=\rho(u_2)=j$ that appear in a bag
together with $v$. Furthermore, the produced decomposition has the property
that each bag contains at least one vertex that does not appear in the previous
bag. \end{lem}

\begin{proof}

Let us start with the last property which states that each bag introduces at
least one new vertex. We can assume that this is always the case, as if
$B_{i+1}\subseteq B_i$, we can delete $B_{i+1}$ from the decomposition and
still have a valid decomposition. Suppose then that we delete a bag that is a
subset of the previous bag, and continue doing this until it is no longer
possible. In the end we obtain a decomposition that satisfies the claimed
property. In the remainder we do not edit this decomposition further, since we
only need to produce a ranking function to complete the proof.

To produce the ranking function, we will consider the interval graph that
corresponds to the given path decomposition and use known facts about the
performance of first-fit coloring on interval graphs. First, suppose that the
given path decomposition has bags $B_1,\ldots, B_{\ell}$. For each $v\in V$, we
denote by $s(v)$ (respectively $t(v)$) the smallest (respectively largest)
index of a bag that contains $v$.  In other words, $v$ appears in all bags in
the interval $[s(v),t(v)]$. It is now well-known (and easy to see) that $G$ is
actually a subgraph of the interval graph $H$ formed by the collection of
intervals $\{ [s(v),t(v)]\ |\ v\in V\}$.  Two vertices are connected in $H$ if
and only if they appear together in a bag of the decomposition. We will produce
the ranking function by appropriately coloring $H$.

In order to produce a ranked path decomposition of $G$, we now need to assign
ranks to vertices of $G$ so that for each $i$, $\rho^{-1}(i)$ forms an independent
set on $H$, as two vertices that have the same rank must not be in the same
bag. We are therefore aiming to color $H$ with at most $8p$ colors, in a way
that respects the extra property of the lemma, which states that each vertex of
rank $i$ must have at most two neighbors of rank $j$ (for $j>i$) in the
interval graph $H$. 

We produce a ranking in the following greedy way: Initially let $S_1=\emptyset$
and as long as there exists $v\in V$ such that $S_1\cup\{v\}$ is an independent
set of $H$, select among such vertices the vertex $v$ that has minimum $t(v)$
and add it to $S_1$.  Once no such vertex $v$ exists (that is, $S_1$ is a
maximal independent set of $H$), we set $\rho(v)=1$ for all $v\in S_1$. We then
remove all vertices of $S_1$ from the graph and recursively execute the same
algorithm on the remaining graph. If $\rho'$ is the ranking returned, we set
for all $v\in V\setminus S_1$ that $\rho(v)=\rho'(v)+1$. In other words, the
idea is that we greedily select an independent set $S_1$, assign its vertices
rank $1$, then greedily select an independent set in the remaining graph and
assign its vertices rank $2$, and so on, until all vertices have an assigned
rank.

It is clear that the ranking we produced respects Definition~\ref{def:rank}, because two
vertices which share a bag are neighbors in $H$ and hence are assigned distinct
ranks. We now prove that no vertex is assigned a rank higher than $8p$, that
is, our coloring uses at most $8p$ colors. To see this, recall the basic
first-fit coloring algorithm, which does the following: given the vertices of a
graph $G$, presented in some order, the algorithm assigns to each vertex the
lowest possible color that does not make any edge, whose other endpoint is
already colored, monochromatic. We can now produce an ordering of $H$ such that
the first-fit algorithm would return a coloring that is identical to our
ranking, by presenting to the algorithm the sets of vertices $\rho^{-1}(1),
\rho^{-1}(2), \ldots$ in this order (with the order inside sets being irrelevant).
It is now not hard to see that, because $\rho^{-1}(1)$ is a maximal independent
set of $G$, the first-fit algorithm will indeed only use color $1$ for
$\rho^{-1}(1)$, and the same can be shown for the remaining colors by induction.
We now invoke a result of \cite{NB08} which states that, on an interval graph
$H$, the first-fit algorithm always produces a coloring with at most $8\chi(H)$
colors, for any vertex ordering.  Since $\chi(H)= \omega(H) = p$ (because $H$
is an interval graph, and interval graphs are perfect), we have that our
ranking uses at most $8p$ ranks.

Finally, the extra property of the ranked path decompositions we produce is
guaranteed by the greedy criterion we used in constructing our ranking. Suppose
that our algorithm sets $\rho(v)=i$ and fix a $j>i$, for $i,j\in[8p]$. We claim
that no vertex $u$ exists that has $\rho(u)=j$ and $[s(u),t(u)]\subseteq [s(v),
t(v)-1]$. If such a vertex $u$ existed, since every neighbor of $u$ is a
neighbor of $v$ in $H$ and $t(u)<t(v)$, then $u$ would have been selected
instead of $v$ when constructing the $i$-th color class. Hence, every vertex
$u$ that is a neighbor of $v$ in $H$ must have either $t(v)\in [s(u),t(u)]$ or
$s(v)\in [s(u),t(u)]$. But, since vertices of rank $j$ are an independent set,
we may have at most one vertex from each case, hence $v$ has at most two
neighbors of rank $j$ in $H$.  \end{proof}

We will call the ranked path decompositions that satisfy the properties of the
decompositions produced by Lemma~\ref{lem:normal} \emph{normalized} path
decompositions. Since such a decomposition can always be obtained without using
too many ranks in the ranking function, we will from now on focus on the case
where we are given a normalized decomposition. Furthermore, we will usually use
$p$ to denote the maximum rank, rather than the pathwidth; this will not have a
significant impact as, according to Lemma~\ref{lem:normal} we can make sure that the
two are at most a constant factor apart.

We note that normalized path decompositions have the useful property that
vertices of rank $1$ appear a bounded number of times, as shown in the
observation below.  Our high-level strategy will be to start from this fact and
edit the graph so that eventually vertices of \emph{any} rank appear a bounded
number of times.

\begin{obs}\label{obs:rankone} Let $G=(V,E)$ be a graph and suppose we
are given a ranked path decomposition of $G$ with ranking function $\rho:V\to
[p]$ satisfying the properties of Lemma~\ref{lem:normal}. Then, for each $u\in V$
with $\rho(u)=1$ we have that $u$ appears in at most $2p$ bags of the
decomposition.  \end{obs}

\begin{proof} Consider such a vertex $u$ with $\rho(u)=1$. Observe that by
Lemma~\ref{lem:normal}, each bag that contains $u$ must introduce a new vertex.
Suppose that $u$ appears in $2p+1$ bags, therefore, there exist at least $2p$
bags that contain $u$ and introduce a new vertex. By pigeonhole principle,
there exists a rank $j\in[2,p]$ such that at least $\frac{2p}{p-1}$ (distinct)
vertices of rank $j$ are introduced in a bag containing $u$. Since
$\frac{2p}{p-1}>2$ this contradicts the assumption that $u$ shares a bag with
at most $2$ distinct vertices of rank $j$.  \end{proof}

\subsection{Finding Isomorphic Bag Intervals}\label{sec:iso}

As mentioned, our high-level strategy will be to identify parts of the graph
which are locally isomorphic, so that we can apply Lemma~\ref{lem:sameN} to obtain a
simpler (less well-connected) graph, and eventually Lemma~\ref{lem:isomorphic} in
order to decrease the size of the graph. In order to identify such parts, we
first define what it means for two blocks of bags of a given decomposition to
be isomorphic.

\begin{defi}\label{def:block-iso} Let $G=(V,E)$ be a $k$-terminal graph
with terminal set $T$, and $B_1,\ldots,B_{\ell}$ a ranked path decomposition of
$G$ with ranking function $\rho:V\to [p]$.  Let $s_1,t_1,s_2,t_2$ be positive
integers with $s_1\le t_1$ and $t_1-s_1=t_2-s_2$. We define the \emph{block}
corresponding to $[s_1,t_1]$ and write $\mathcal{B}(s_1,t_1)$ to be $\{B_j\ |\
j\in[s_1,t_1]\ \}$. We say that $\mathcal{B}(s_1,t_1)$ is
\emph{block-isomorphic} to $\mathcal{B}(s_2,t_2)$ if

\begin{enumerate}

\item For each $j\in [s_1,t_1]$ and rank $i$ we have $|\rho^{-1}(i)\cap
B_j|=|\rho^{-1}(i)\cap B_{s_2+(j-s_1)}|$.

\item For each $j\in [s_1+1,t_1]$ and rank $i$ we have that $B_j$ contains a
vertex $v$ with $\rho(v)=i$ such that $v\not\in B_{j-1}$ if and only if
$B_{s_2+(j-s_1)}$ contains a vertex $v'$ with $\rho(v')=i$ such that $v'\not\in
B_{s_2+(j-s_1)-1}$.

\item The following mapping $f$ is a terminal-respecting isomorphism from
$G[T\cup(\bigcup_{j\in[s_1,t_1]}B_j)]$ to
$G[T\cup(\bigcup_{j\in[s_2,t_2]}B_j)]$.  For each $v\in
\bigcup_{j\in[s_1,t_1]}B_j$ we let $j_v$ be the minimum index in $[s_1,t_1]$
such that $v\in B_{j_v}$ and define $f(v)$ to be the (unique) vertex of
$B_{s_2+(j_v-s_1)}$ such that $\rho(v)=\rho(f(v))$.

\end{enumerate} 

\end{defi}

\begin{defi} Let $L\ge 0$ and $G,k,T,\ell,\rho$ as in
Definition~\ref{def:block-iso}.  For positive integers $s_1,s_2\in [\ell-L]$ we will
write $s_1\approx_L s_2$ to indicate that $\mathcal{B}(s_1,s_1+L)$ is
block-isomorphic to $\mathcal{B}(s_2,s_2+L)$.  \end{defi}

Note that the isomorphism of Definition~\ref{def:block-iso} is well-defined, because
according to the first condition, if $B_j$ contains a vertex of rank $i$, then
so does $B_{s_2+(j_v-s_1)}$, and such a vertex is unique by the definition of
ranked path decomposition. According to Definition~\ref{def:block-iso}, two blocks of
bags are isomorphic only if the subgraphs induced by the bags they contain (and
the terminals of $G$) are isomorphic under the trivial mapping function which
maps each vertex of a bag from one block to the vertex of the corresponding bag
of the other block that has the same rank. Despite the fact that this restricts
the class of isomorphisms we may consider quite a bit, the block-isomorphism
relation is an equivalence relation that does not have too many equivalence
classes.  In particular, we have the following.

\begin{lem} Let $L\ge 0$, $G=(V,E)$ be a $k$-terminal graph with terminal set
$T$, and $B_1,\ldots,B_{\ell}$ a ranked path decomposition of $G$ with ranking
function $\rho:V\to [p]$. Let $t_1,t_2$ be positive integers such that for all
$j,j'\in[t_1,t_2]$ we have $B_j\cap T=B_{j'}\cap T$.  Then, the relation
$\approx_L$ is an equivalence relation on the set $[t_1,t_2-L]$ with at most
$2^{(L+1)(p^2+2p+kp)}$ equivalence classes.  \end{lem}

\begin{proof} 

The fact that $\approx_L$ is an equivalence relation is easy to see, as
terminal-respecting isomorphisms can be composed to show transitivity. The
interesting part of the lemma is then the bound on the number of equivalence
classes. We prove this by induction on $L$. 

For $L=0$, we claim there are at most $2^{p+p^2+kp}$ equivalence classes of
bags (in this case, each block consists of a single bag). Indeed, in order to
decide if $\mathcal{B}(s_1,s_1)=\{B_{s_1}\}$ and
$\mathcal{B}(s_2,s_2)=\{B_{s_2}\}$ are block-isomorphic, we first need to check
if $B_{s_1},B_{s_2}$ contain vertices of the same ranks, and for this there are
$2^p$ equivalence classes. If they do, then we must check, for each $i_1,i_2\in
[p]$ if the vertices of ranks $i_1,i_2$ in each of $B_{s_1}, B_{s_2}$ are
adjacent, and for this we have $2^{p\choose 2}<2^{p^2}$ equivalence classes.
Finally, since the isomorphism has to be terminal-respecting, we have to check
for each rank $i\in [p]$ if the vertex of rank $i$ in each of $B_{s_1},
B_{s_2}$ is connected to each of the $k$ terminals, which gives at most $kp$
edges which may or may not exist. (Note that we have to check these edges, even
though the two bags contain the same terminals, because terminal-respecting
isomorphisms must also preserve the edges between terminals outside the bag).
Overall we have at most $2^{p+p^2+kp}<2^{p^2+2p+kp}$ choices.  If we make the
same choices for two bags, the two bags are block-isomorphic, hence we have
bounded the number of equivalence classes for $L=0$.

Suppose now that $L>0$ and we have shown that the number of equivalence classes
of $\approx_{L-1}$ is at most $2^{L(p^2+2p+kp)}$. Consider two indices
$s_1,s_2$ for which we want to check if $s_1\approx_L s_2$. We claim that for
this it is sufficient to have $s_1\approx_{L-1} s_2$ and to satisfy certain
conditions for the bags $B_{s_1+L}, B_{s_2+L}$ for which we have at most
$2^{p^2+2p+kp}$ choices. More precisely, for each rank $i$, we have three
possibilities for the bag $B_{s_1+L}$: either the bag contains no vertex of
rank $i$; or it contains a vertex of rank $i$ that also appears in
$B_{s_1+L-1}$; or it contains a vertex of rank $i$ that appears for the first
time in $B_{s_1+L}$ (and hence this vertex is a non-terminal). Suppose now that
for each rank $i$, the bags $B_{s_1+L}, B_{s_2+L}$ agree on the choice of which
of these three possibilities holds (there are $3^p<2^{2p}$ possibilities in
total), and furthermore, that the graphs induced by $B_{s_1+L}\cup T$ and
$B_{s_2+L}\cup T$ are isomorphic for the natural terminal-respecting
isomorphism that matches vertices of the same rank (at most $2^{p^2+kp}$
possibilities). Then, if $s_1\approx_{L-1}s_2$, we now have $s_1\approx_L s_2$.
Therefore, each of the $2^{L(p^2+2p+kp)}$ equivalence classes of
$\approx_{L-1}$ has been refined into at most $2^{2p+p^2+kp}$ equivalence
classes, giving that the number of equivalence classes of $\approx_L$ is at
most $2^{(L+1)(p^2+2p+kp)}$, as desired.  \end{proof}

Now that we know that block-isomorphism has a bounded number of equivalence
classes (if $k,p,L$ are bounded), we can try to look for ``copies'' of the same
block in our path decomposition. We observe the following lemma.

\begin{lem}\label{lem:php1} Let $L$ be a non-negative integer, $G=(V,E)$ be a
$k$-terminal graph with terminal set $T$, and $B_1,\ldots,B_{\ell}$ a ranked
path decomposition of $G$ with ranking function $\rho:V\to [p]$. We define
$R=(L+1)(2^{(L+1)(p^2+2p+kp)}+1)$. Let $t_1,t_2$ be positive integers with
$t_2\ge t_1+R$ such that for all $j,j'\in[t_1,t_2]$ we have $B_j\cap
T=B_{j'}\cap T$.  Then, for every $s\in[t_1,t_2-R]$, there exist $s_1,s_2\in
[s,s+R-(L+1)]$ such that $s_1+L<s_2$ and $s_1\approx_L s_2$.  \end{lem}

\begin{proof} Consider the set of indices $S=\{s, s+L+1, s+2(L+1), \ldots,
s+R-(L+1)\}$. The number of distinct indices in $S$ is
$\frac{R}{(L+1)}=2^{(L+1)(p^2+2p+kp)}+1$. Therefore, by pigeonhole principle
there must exist $s_1,s_2\in S$ such that $s_1\approx_L s_2$, as $\approx_L$
has $2^{(L+1)(p^2+2p+kp)}$ equivalence classes. Since any two elements of $S$
have difference at least $L+1$, the lemma follows.  \end{proof}

What we have shown so far is that if we take sufficiently many (at least $R$)
consecutive bags in our decomposition, we will find two blocks of length
(roughly) $L$ which are block-isomorphic. Let us now move a step further.

\begin{lem}\label{lem:php2} Let $L$ be a non-negative integer, $G=(V,E)$ be a
$k$-terminal graph with terminal set $T$, $B_1,\ldots,B_{\ell}$ a ranked path
decomposition of $G$ with ranking function $\rho:V\to [p]$. Let $q,R$ be
positive integers. We define $R^*=(R+1) (q2^{(R+1)(p^2+2p+kp)}+1)$. Let
$t_1,t_2$ be positive integers with $t_2\ge t_1+R^*$. Then, for every
$s\in[t_1,t_2-R^*]$ there exist $q+1$ distinct $s_1,s_2,\ldots, s_{q+1}\in
[s,s+R^*-(R+1)]$, such that for any two distinct $s_i,s_j$ with $i,j\in [q+1]$
we have $|s_i-s_j|>R$ and $s_i\approx_R s_j$.  \end{lem}

\begin{proof} As in the proof of Lemma~\ref{lem:php1}, consider the set of indices
$S=\{s, s+R+1, s+2(R+1),\ldots, s+R^*-(R+1)\}$. These are
$\frac{R^*}{R+1}=q2^{(R+1)(p^2+2p+kp)}+1$ distinct indices, all of which have
pairwise difference at least $R+1$. Since $\approx_R$ has
$2^{(R+1)(p^2+2p+kp)}$ equivalence classes, by pigeonhole principle there must
exist a subset of $S$ of size at least $q+1$ such that any two of its elements
are equivalent for $\approx_R$.  \end{proof}

Note that Lemma~\ref{lem:php1} and Lemma~\ref{lem:php2} are non-vacuous only if we find a
long enough interval where all bags contain the same terminals, that is if
$t_2-t_1\ge R$ or $t_2-t_1\ge R^*$ respectively. We will take this into account
when we use these lemmas in the next section.

At this point we are almost done in our search for appropriate isomorphic parts
of the graph. What we have proved is that, if we fix some appropriate radius
$L$, there is some larger radius $R^*$ (double-exponential in $L$), such that
if we look at any interval of the path decomposition of length $R^*$, we will
be able to find $q+1$ isomorphic $R$-blocks, which are long enough to guarantee
the existence of two isomorphic $L$-blocks inside them. What remains is to ask
what value of $L$ will be appropriate for our purposes. Ideally, we would like
to calculate a value $L$ that will allow us to preserve the balls around
vertices for a suitable radius and apply Lemma~\ref{lem:sameN}. However, we can only
give such a bound if we know that vertices of our path decomposition do not
appear in too many bags.

\begin{lem}\label{lem:pwD} Let $G=(V,E)$ be a $k$-terminal graph with
terminal set $T$ and $B_1,\ldots,B_{\ell}$ a ranked path decomposition of $G$
with the additional property that any non-terminal vertex appears in at most
$\Delta$ bags of the decomposition. Then, for each $r\ge 0$ and for each
non-terminal vertex $v$, if $v\in B_j$, then each non-terminal vertex of
$B_r(v)$ is contained in a bag of $\mathcal{B}(j-r\Delta,j+r\Delta)$.
\end{lem}

\begin{proof}

We prove the lemma by induction on $r$. If $r=0$, then the only non-terminal
vertex of $B_r(v)$ is $v$ itself, which is contained in $B_j\in
\mathcal{B}(j,j)$.

Suppose $r>0$ and the lemma is true up to $r-1$. Consider a non-terminal vertex
$u$ of $B_r(v)$. If there is a path without internal terminal vertices from $u$
to $v$ that has length at most $r-1$, then by inductive hypothesis $u$ is
contained in a bag of $\mathcal{B}(j-(r-1)\Delta,j+(r-1)\Delta)$, hence also in
$\mathcal{B}(j-r\Delta,j+r\Delta)$. Suppose then that the shortest such path
from $u$ to $v$ has length $r$, and that the vertex immediately after $u$ in
this path is $w$. By inductive hypothesis, since $w$ is at distance at most
$r-1$ from $v$, it is contained in a bag of
$\mathcal{B}(j-(r-1)\Delta,j+(r-1)\Delta)$. Furthermore, $w$ must be contained
in a bag together with $u$, since they are neighbors. But, if $u$ does not
appear in any bag with index in the interval $[j-r\Delta,j+r\Delta]$, that
means that $w$ must appear either in $B_{j-r\Delta-1}$ or in $B_{j+r\Delta+1}$.
In both cases, $w$ appears in strictly more than $\Delta$ bags, contradiction.
Therefore, $u$ must appear in some bag with index in the interval
$[j-r\Delta,j+r\Delta]$, as desired. \end{proof}

\subsection{Rewiring Operation}\label{sec:rewiring}

The goal of Section~\ref{sec:iso} was to present the basic tools which will allow us
to find isomorphic parts of the input graph. Ideally, we would then like to use
Lemma~\ref{lem:isomorphic} and delete one such part. However, this is in general not
possible, as the isomorphism guaranteed by the lemmas of Section~\ref{sec:iso} is not
sufficient to obtain identical sets, in the sense of Definition~\ref{def:id}. What we
need to do, then, is to edit the graph in a way that does not affect the
validity of any FO formula with $q$ quantifiers but leverages the isomorphic
parts we have found to construct $q+1$ identical parts on which
Lemma~\ref{lem:isomorphic} can be applied. We now present the basic edit operation
which will allow us to achieve this for appropriate parameters.

\begin{defi}[Rewiring]\label{def:rewiring} Let $G=(V,E)$ be a
$k$-terminal graph for which we are given a ranked path decomposition with
ranking function $\rho:V\to[p]$.  Let $B_{s_1}, B_{s_2}$ be two bags of this
decomposition, for $s_1<s_2$. We define the \emph{rewiring} operation on
$(s_1,s_2)$ as follows: 

\begin{enumerate}

\item For every non-terminal vertex $v\in B_{s_1}$ which is adjacent to a
non-terminal vertex $u\in B_j\setminus (B_{s_1}\cup B_{s_2})$ for some $j\in
[s_1+1,s_2-1]$ we delete the edge $uv$ and add to the graph the edge $uv'$,
where $v'\in B_{s_2}$ and $\rho(v)=\rho(v')$, if such a $v'$ exists 

\item For every non-terminal vertex $v\in B_{s_2}$ which is adjacent to a
non-terminal vertex $u\in B_j\setminus B_{s_2}$ for some $j>s_2$, we delete the
edge $uv$ and add to the graph the edge $uv'$, where $v'\in B_{s_1}$ and
$\rho(v)=\rho(v')$, if such a $v'$ exists.  

\end{enumerate}

\end{defi}

\begin{figure} \includegraphics{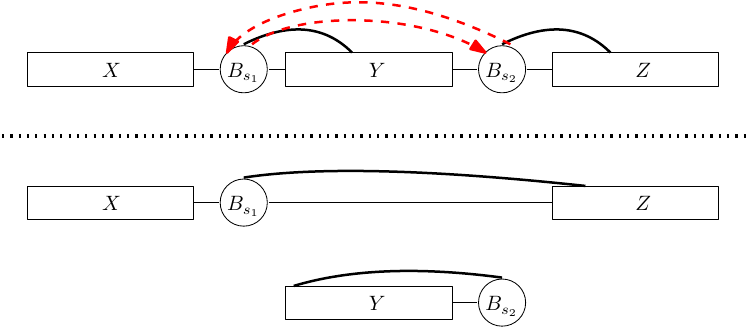}\caption{An illustration of the
rewiring operation of Definition~\ref{def:rewiring}. Edges from $Y$ to $B_{s_1}$ are
rerouted towards $B_{s_2}$, while edges from $Z$ to $B_{s_2}$ are rerouted
towards $B_{s_1}$. Note that edges incident to terminals are not
modified.}\label{fig:rewiring} \end{figure}

Before we proceed, some explanations are in order regarding the motivation of
the rewiring operation of Definition~\ref{def:rewiring}. We first invite the reader to
take a look at Figure~\ref{fig:rewiring}. Recall that from standard properties of
path decompositions, $B_{s_1}, B_{s_2}$ are separators which break down the
graph into three parts, call them $X,Y,Z$, which are respectively vertices
which appear in a bag before $B_{s_1}$, between $B_{s_1}$ and $B_{s_2}$, and
after $B_{s_2}$. The rewiring operation leaves all edges incident to terminals
and all edges incident to $X$ unchanged. What it does is replace edges from $Y$
to $B_{s_1}$ with edges from $Y$ to $B_{s_2}$ and edges from $Z$ to $B_{s_2}$
with edges from $Z$ to $B_{s_1}$.  Intuitively, what this is meant to achieve
is to break down the long path-like structure $X-B_{s_1}-Y-B_{s_2}-Z$ into the
shorter path-like structure $X-B_{s_1}-Z$ and the ring-like structure
$Y-B_{s_2}$. The idea here is that the $Y-B_{s_2}$ part is ``disconnected''
from the rest of the graph (more precisely, the $k$ terminals separate this
part from the rest of the graph, since terminals are not modified by this
operation). Hence if we find many isomorphic such parts, they will also be
identical in the sense of Definition~\ref{def:id}, allowing us to delete one using
Lemma~\ref{lem:isomorphic}. This argument is made precise in Lemma~\ref{lem:foundid}. We
also note that the ``if such a $v'$ exists'' part of the operation is only
added to make the operation well-defined in all cases; we will in fact only
apply this operation in a context where $v'$ is guaranteed to exist.

Before we do all these things, however, we need to be sure that the rewiring
operation did not affect the validity of any FO formula of at most $q$
quantifiers. The main claim now is that if $s_1,s_2$ are sufficiently far
apart, we have a bound on the number of occurrences of non-terminal vertices in
bags, and a sufficiently large block around $B_{s_1}$ is block-isomorphic to a
sufficiently large block around $B_{s_2}$, then the ball of radius $r=2^q-1$
around any vertex has remained unchanged. Hence, we can invoke Lemma~\ref{lem:sameN}
to conclude that the rewired graph is indistinguishable from the original graph
for FO formulas with $q$ quantifiers. Our main tool in proving this will be the
following lemma.

\begin{lem}\label{lem:rewire} Let $G=(V,E)$ be a $k$-terminal graph with
terminal set $T$, $B_1,\ldots,B_{\ell}$ a ranked path decomposition of $G$ with
ranking function $\rho:V\to [p]$ with the additional property that any
non-terminal vertex appears in at most $\Delta$ bags of the decomposition.  Fix
an integer $q\ge 0$ and let $L=\Delta(2^q-1)$. Let $s_1,s_2$ be such that (i)
we have $s_1>4L$, $s_2<\ell-4L$, $s_2-s_1>6L$ (ii) $\mathcal{B}(s_1-L,s_1+L)$
is block-isomorphic to $\mathcal{B}(s_2-L,s_2+L)$.  Let $G'$ be the graph
obtained by applying the rewiring operation on $(s_1,s_2)$. Then, for all FO
formulas $\phi$ with at most $q$ quantifiers we have $G\models \phi$ if and
only if $G'\models \phi$.  \end{lem}

\begin{figure} \includegraphics{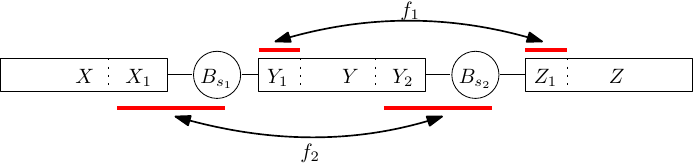}\caption{Schematic view of the two
mappings of the proof of Lemma~\ref{lem:rewire}.}\label{fig:rewire2}\end{figure}

\begin{proof}

Note that $G,G'$ are two graphs on the same set of vertices, for which we use
the same labeling function (the rewiring operation only edits edges of the
graph).  Our plan is to use the identity mapping $f:V\to V$ which sets $f(v)=v$
for all $v\in V$ and prove that this mapping is sufficient to invoke
Lemma~\ref{lem:sameN}.  The mapping is clearly bijective and maps terminals to
themselves, hence the first condition is trivially satisfied.  What remains is
to prove that for all non-terminal vertices $v$, $B_r^{G}(v)$ is isomorphic to
$B_r^{G'}(v)$, where $r=2^q-1$.

Let us identify some interesting sets of vertices of $G$ (we refer to
Figure~\ref{fig:rewire2}).  We define the following:

\begin{enumerate}

\item $X_1$ is the set of all non-terminal vertices which appear in a bag
$B_j$, with $j\in [s_1-L,s_1-1]$ but do not appear in $B_{s_1}$.

\item $Y_1$ is the set of all non-terminal vertices which appear in a bag
$B_j$, with $j\in [s_1+1,s_1+L]$ but do not appear in $B_{s_1}\cup B_{s_2}$.  

\item $Y_2$ is the set of all non-terminal vertices which appear in a bag
$B_j$, with $j\in [s_2-L,s_2-1]$ but do not appear in $B_{s_1}\cup B_{s_2}$.  

\item $Z_1$ is the set of all non-terminal vertices which appear in a bag
$B_j$, with $j\in [s_2+1,s_2+L]$ but do not appear in $B_{s_2}$.  

\end{enumerate}

Recall that since $\mathcal{B}(s_1-L,s_1+L)$ is block-isomorphic to
$\mathcal{B}(s_2-L,s_2+L)$, there exists a terminal-respecting isomorphism
$f_*$ from $G[T\cup(\bigcup_{j\in[s_1-L,s_1+L]}B_j)]$ to
$G[T\cup(\bigcup_{j\in[s_2-L,s_2+L]}B_j)]$ as prescribed in
Definition~\ref{def:block-iso}.  We claim that $f_*$ is a bijection from $X_1$ to $Y_2$,
from $B_{s_1}$ to $B_{s_2}$, and from $Y_1$ to $Z_1$. This follows immediately
from Definition~\ref{def:block-iso}, because each vertex $v\in B_{s_1-L}$ must be mapped
to a vertex of $B_{s_2-L}$, and each vertex that is introduced in $B_j$ for
$j\in [s_1-L+1,s_1+L]$ must be mapped to a vertex of the bag $B_{s_2+(j-s_1)}$.

We use $f_*$ to construct some helpful mappings as follows:

\begin{enumerate}

\item $f_1:V\to V$ is defined as $f_1(v) = \left\{\begin{array}{ll}  f_*(v) &
\textrm{if }v\in Y_1 \\ f_*^{-1}(v) & \textrm{if }v\in Z_1 \\ v
&\textrm{otherwise} \end{array} \right.$

\item $f_2:V\to V$ is defined as $f_2(v) = \left\{\begin{array}{ll}  f_*(v) &
\textrm{if }v\in B_{s_1}\cup X_1 \\ f_*^{-1}(v) & \textrm{if }v\in B_{s_2}\cup
Y_2 \\ v &\textrm{otherwise} \end{array} \right.$

\end{enumerate}

We observe that $f_1,f_2$ are bijections from $V$ to $V$. In particular, $f_1$
translates $Y_1$ to $Z_1$, using the bijective mapping $f_*$ between them, and
leaves all other vertices unchanged, while $f_2$ translates $B_{s_1}\cup X_1$
to $B_{s_2}\cup Y_2$. We invite the reader to take a look at Figure~\ref{fig:rewire2}
for a schematic view of these mappings.

We now claim that for any non-terminal vertex $v$, either $f_1$ or $f_2$ is a
terminal-respecting isomorphism from $B_r^{G}(v)$ to $B_r^{G'}(v)$. In
particular, we have two cases:

\begin{enumerate}

\item If $v$ appears in a bag $B_j$ with $j\in [1,s_1]\cup [s_2-3L,s_2]$, then
$f_1$ is a terminal-respecting isomorphism from $B_r^{G}(v)$ to $B_r^{G'}(v)$.

\item Otherwise, that is, if $v$ only appears in bags $B_j$ with $j\in
[s_1+1,s_2-3L-1]\cup [s_2+1,\ell]$, then $f_2$ is a terminal-respecting
isomorphism from $B_r^{G}(v)$ to $B_r^{G'}(v)$.

\end{enumerate}

Indeed, suppose that $v$ appears in a bag $B_j$ with $j\in[1,s_1]$. Then,
according to Lemma~\ref{lem:pwD}, all vertices of $B_r^G(v)$ appear in a bag $B_j$
with $j\in[1,s_1+L]$. All such vertices are mapped to themselves by $f_1$,
except for vertices of $Y_1$ which are mapped to $Z_1$ in a way that preserves
internal edges. What we need, then, is to argue that the rewiring operation
translates edges from $Y_1$ to $B_{s_1}$ to edges from $Z_1$ to $B_{s_1}$ (we
do not need to worry about edges from $Y_1$ to $B_j$ for $j<s_1$, as $B_{s_1}$
is a separator).  Now we claim that for every $y\in Y_1$ which had an edge to
$v\in B_{s_1}$ in $G$, the vertex $f_*(y)\in Z_1$ has an edge to $v$ in $G'$,
by the rewiring operation. More precisely, the edge $yv$ has been deleted by
the first part of the rewiring operation, but because $f_*$ is an isomorphism,
we know that $f_*(y)f_*(v)$ is an edge of $G$, with $f_*(v)\in B_{s_2}$, and
the second part of the rewiring operation will replace this edge with the edge
$f_*(y)v$.

Suppose then that $v$ appears in a bag $B_j$ with $j\in[s_2-3L,s_2]$, therefore
all vertices of $B_r^G(v)$ appear in a bag $B_j$ with index
$j\in[s_2-4L,s_2+L]$. All such vertices are mapped to themselves by $f_1$,
except for vertices of $Z_1$, which are mapped to $Y_1$. As in the previous
paragraph, we need to check that if for $z\in Z_1$ and $v\in B_{s_2}$ we have
the edge $zv$ in $G$, then we have the edge $f_*^{-1}(z)v$ in $G'$. Because
$f_*$ is an isomorphism, we know that $f_*^{-1}(z)f_*^{-1}(v)$ is an edge of
$G$, with $f_*^{-1}(z)\in Y_1$ and $f_*^{-1}(v)\in B_{s_1}$. The first part of
the rewiring operation will replace the edge $f_*^{-1}(z)f_*^{-1}(v)$ with
$f_*^{-1}(z)v$.

Consider now the case where $v$ appears in a bag $B_j$ with $j\in
[s_1+1,s_2-3L-1]$, therefore all vertices of $B_r^G(v)$ are contained in a bag
$B_j$ with $j\in [s_1+1-L,s_2-2L-1]$. All such vertices are mapped to
themselves by $f_2$ except vertices of $X_1\cup B_{s_1}$, which are mapped to
$Y_2\cup B_{s_2}$. What we need to argue now is that the rewiring operation has
translated edges from $Y_1$ to $B_{s_1}$ to edges from $Y_1$ to $B_{s_2}$
(edges from $Y_1$ to $X_1$ don't exist, as $B_{s_1}$ is a separator). Let $y\in
Y_1$ and $v\in B_{s_1}$ such that $yv$ is an edge of $G$. The first part of the
rewiring operation will replace this edge with $yf_*(v)$, with $f_*(v)\in
B_{s_2}$.

Finally, if $v$ appears in a bag $B_j$ with $j\in [s_2,\ell]$, we have that all
vertices of $B_r^G(v)$ are contained in a bag with index in $[s_2-L,\ell]$ and
$f_2$ maps all such vertices to themselves, except vertices of $Y_2\cup
B_{s_2}$ are mapped to $X_1\cup B_{s_1}$. We need to check that an edge from
$Z_1$ to $B_{s_2}$ is mapped to an edge from $Z_1$ to $B_{s_1}$ in $G'$.
Suppose that for $z\in Z_1$ and $v\in B_{z_2}$ we have the edge $zv$ in $G$.
The second part of the rewiring operation will replace this edge with the edge
$zf_*^{-1}(v)$, with $f_*^{-1}(v)\in B_{s_1}$.

We have now explained how for each non-terminal vertex $v$, $B_r^G(v)$ and
$B_r^{G'}(v)$ are isomorphic, thanks to either $f_1$ or $f_2$. The only thing
that remains is to explain why these mappings are also terminal-respecting.
However, this is easy to see as these mappings either map a vertex to itself,
or map $v$ to $f_*(v)$ (or $f_*^{-1}(v)$), and $f_*(v)$ is a
terminal-respecting isomorphism. As a result, if $v$ is connected to a terminal
vertex, then $f_1(v), f_2(v)$ are also connected to the same terminal vertex
and our mapping is terminal-respecting.  \end{proof}

Finally, what we need to argue to complete this section is that if we apply the
rewiring operation on two block-isomorphic parts of the graph, then we will
obtain two parts of the graph which are identical in the sense of
Definition~\ref{def:id}. This will allow us to delete a part of the graph, once we
accumulate a large enough number of identical parts.

\begin{lem}\label{lem:foundid} Let $R$ be a positive integer, $G=(V,E)$ be a
$k$-terminal graph with terminal set $T$, and $B_1,\ldots,B_{\ell}$ a ranked
path decomposition of $G$ with ranking function $\rho:V\to [p]$ with the
property that no non-terminal vertex appears in more than $R$ bags.  Let
$s_1,s_2$ be positive integers such that $s_2-s_1>4R$ and
$\mathcal{B}(s_1,s_1+R)$ is block-isomorphic to $\mathcal{B}(s_2,s_2+R)$.  Let
$j_1,j_2\in [0,R-1]$ with $j_1<j_2$ and let $G'$ be the graph obtained after
applying the rewiring operation on $(s_1+j_1,s_1+j_2)$ and also on
$(s_2+j_1,s_2+j_2)$. Let $Y_1$ be the set of vertices that appear in a bag with
index in $[s_1+j_1+1,s_1+j_2-1]$, but not in $B_{s_1+j_1}\cup B_{s_1+j_2}$.
Similarly, let $Y_2$ be the set of vertices that appear in a bag with index in
$[s_2+j_1+1,s_2+j_2-1]$, but not in $B_{s_2+j_1}\cup B_{s_2+j_2}$. Then
$(Y_1\cup B_{s_1+j_2})\setminus T$ is identical to $(Y_2\cup
B_{s_2+j_2})\setminus T$. \end{lem}

\begin{proof}

First, observe that $s_1,s_2$ are far enough apart that the two rewiring
operations are applied to disjoint sets of non-terminals (and terminal vertices
are unaffected by these operations), so it is not important which operation
happens ``first''.

Since $\mathcal{B}(s_1,s_1+R)$ is block-isomorphic to $\mathcal{B}(s_2,s_2+R)$
there is a terminal-respecting isomorphism $f_*$ which maps $Y_1\cup
B_{s_1+j_2}$ to $Y_2\cup B_{s_2+j_2}$. More precisely, $f_*$ maps vertices of
$\bigcup_{j\in [s_1+j_1+1,s_1+j_2]} B_j$ to $\bigcup_{j\in [s_2+j_1+1,s_2+j_2]}
B_j$, and by excluding vertices of $B_{s_1+j_1}$ we get a mapping from $Y_1\cup
B_{s_1+j_2}$ to $Y_2\cup B_{s_2+j_2}$. It is easy to see that if $f^*$ is an
isomorphism before we apply the rewiring operation, it remains an isomorphism
afterwards, since we apply the same operation on isomorphic graphs.

We claim that this isomorphism also proves that the two sets are identical.
Indeed, consider the mapping that maps each vertex  $v\in Y_1\cup B_{s_1+j_2}$
to $f_*(v)$, maps each vertex $v\in Y_2\cup B_{s_2+j_2}$ to $f_*^{-1}(v)$, and
maps all other vertices to themselves. To see that this is a
terminal-respecting isomorphism, we observe that adjacency relations with
terminals are preserved, as $f_*$ is terminal-respecting. To see that edges
between non-terminals are preserved, we note that edges with both endpoints
inside $Y_1\cup B_{s_1+j_2}$ or $Y_2\cup B_{s_2+j_2}$ are preserved because
$f_*$ is an isomorphism, while edges with both endpoints outside the sets are
trivially preserved. What remains is to note that there is no non-terminal in
$Y_1\cup B_{s_1+j_2}\cup Y_2\cup B_{s_2+j_2}$ which has an edge to a
non-terminal outside the set. This is a result of the rewiring operation: every
edge from $Y_1$ to $B_{s_1+j_1}$ has been replaced by an edge to $B_{s_1+j_2}$,
and since $B_{s_1+j_1}$ is a separator, there are no edges from $Y_1$ towards
bags further to the left. Similarly, every edge from $B_{s_1+j_2}$ to a vertex
that appears in $B_{s_1+j_2+1}$ or later has been replaced with an edge from
$B_{s_1+j_1}$. Hence, the sets $Y_1\cup B_{s_1+j_2}, Y_2\cup B_{s_2+j_2}$ have
no edges connecting them to the outside without going through terminals and the
mapping we defined proves they are identical.  \end{proof}

\section{Putting Everything Together}\label{sec:final}

We are now finally ready to put everything together to obtain our model
checking algorithm for FO logic. Our plan of attack is to formulate a procedure
which can either simplify the graph in a way that does not affect the validity
of the given formula (or any formula with the same number of quantifiers), or
we can certify that the graph has bounded degree, and hence we can use known
model checking procedures with an elementary dependence on the formula. On a
high level, we take as input a graph $G$, a path decomposition of $G$, and a
formula $\phi$ with $q$ quantifiers and we will do the following:

\begin{enumerate}

\item Use Lemma~\ref{lem:normal} to normalize the decomposition and obtain a ranking
of the vertices. Let $p$ be the number of ranks used in the resulting ranking
(which per Lemma~\ref{lem:normal} is linear in the width of the original
decomposition). In this ranking, vertices of rank $1$ appear in a constant
number of bags (Observation~\ref{obs:rankone}).  We would like to extend this so that
every vertex appears in a bounded number of bags.  In the remainder we will use
the number of times a vertex appears in the path decomposition as a proxy bound
for its degree.

\item Define a bound function $\Delta(i)$ which defines an acceptable bound for
the number of occurrences in distinct bags for a vertex of rank $i$. This
function will be a tower of exponentials of height roughly $2i$, but this is
acceptable, since the maximum rank is upper-bounded by a function of the
pathwidth, which we consider to be an absolute constant.

\item Examine the graph and check if any vertex of rank $i$ appears in more
than $\Delta(i)$ bags. If this is not the case, we can bound the maximum degree
of the graph, and we are done.

\item Otherwise, find a vertex $v$ of minimum rank $i$ that appears more than
$\Delta(i)$ times. Find a section of the decomposition where $v$ appears, and
where all bags contain the same vertices of rank higher than $i$ (if
$\Delta(i)$ is large, we can find such a section that is quite long).  We label
as terminals the vertices of the first and last bag of the section, and the
vertices of rank at least $i$ appearing in the section. 

\item Now, the remaining vertices of the section appear a bounded (by
$\Delta(i-1)$) number of times, and are separated by the rest of the graph by
$k=O(p)$ terminals. However, they are quite numerous, as we assumed that $v$
appears too many times. Therefore, we can invoke the machinery of
Section~\ref{sec:iso} to find some isomorphic parts. Note that it is important that
vertices of rank at least $i$ (which are now terminals) are common throughout
the section, which allows us to invoke Lemmas \ref{lem:php1} and
\ref{lem:php2}.

\item Having found many isomorphic parts, we use the tools of
Section~\ref{sec:rewiring} to perform the rewiring operation that will produce $q+1$
identical parts, of which we can remove one. We then ``undo'' the operation on
the remaining parts, and obtain a smaller graph, where $v$ appears in fewer
bags, without changing whether $\phi$ is satisfied.

\end{enumerate}

\begin{defi}\label{def:delta} Let $p,q$ be positive integers. We define
the function $\Delta_{p,q}(i)$ as follows: $\Delta_{p,q}(1)=5p$ and
$\Delta_{p,q}(i+1)=2^{2^{\Delta(i)\cdot 2^{20qp^2}}}$. When $p,q$ are clear
from the context, we will write $\Delta(i)$ to denote $\Delta_{p,q}(i)$.
\end{defi}

\begin{obs} For each fixed $p,i$, the function $\Delta_{p,q}(i)$ is an
elementary function of $q$. Furthermore, $\Delta_{p,q}(i)$ is a strictly
increasing function of $i$.\end{obs}

\begin{lem}\label{lem:main} Fix two positive integers $p,q$. There is an
algorithm that takes as input an $n$-vertex graph $G=(V,E)$, a normalized
ranked path decomposition of $G$ with ranking function $\rho:V\to [p]$, such
that $G$ contains a vertex that appears in at least $5p\cdot \Delta(p)$ bags of
the decomposition.  Then, the algorithm runs in time polynomial in $n$ and
outputs a smaller graph $G'$ and a normalized ranked path decomposition of $G'$
with the same ranking function $\rho$, such that for all FO formulas $\phi$
with at most $q$ quantifiers we have $G\models \phi$ if and only if $G'\models
\phi$.  \end{lem}

\begin{proof}

Fix an integer $q$. We will describe an algorithm that takes the input
mentioned in the lemma and outputs a graph $G'$. Fix some arbitrary FO formula
$\phi$ with at most $q$ quantifiers and we want to establish that $G\models
\phi$ if and only if $G' \models \phi$. 

We begin by locating a vertex $v$ such that $\rho(v)=i$, $v$ appears in at
least $5p\cdot\Delta(i)$ bags, and $i$ is minimized. Note that such a vertex exists,
since we assume that there exists a vertex which appears $5p\cdot\Delta(p)$ times.
We now have the property that any vertex of rank at most $i-1$ appears at most
$5p\cdot\Delta(i-1)$ times, since $i$ was selected to be minimum. Note that
$i>1$, because by Lemma~\ref{lem:normal} and Observation~\ref{obs:rankone}, vertices of rank
$1$ appear at most $2p$ times. 

Suppose that $v$ appears in the bags $B_j$ with indices in the interval
$[a_1,a_2]$.  We say that a bag $B_j$ of this interval is interesting, if there
exists $i'>i$ such that a vertex $v'$ with $\rho(v')=i'$ belongs in exactly one
of $B_j, B_{j-1}$.  In other words, a bag is interesting if the higher-rank
vertices it contains changed with respect to the previous bag. We observe that
there are at most $4p$ interesting bags, because $v$ shares a bag with at most
$2p$ vertices of higher rank, and an interesting bag can either forget (remove)
a higher-rank vertex which was already present (once), or add a higher-rank
vertex (once). 

Define a section to be a maximal contiguous interval of bags which contain the
same vertices of rank at least $i$. In the interval $[a_1,a_2]$ every bag
contains $v$ as its unique vertex of rank $i$, so a section in this interval
must start with an interesting bag, except perhaps the section starting at
$a_1$. There are therefore at most $4p+1$ sections of bags that contain $v$.
Therefore, there must exist a section that contains at least
$\frac{5p\Delta(i)}{4p+1}\ge \Delta(i)$ bags, suppose this section is
$[a_3,a_4]\subseteq [a_1,a_2]$.  All bags of the section contain the same
vertices of rank $i'\ge i$.  We now add a labeling function $\mathcal{T}$ to
$G$ and set as terminals all the vertices of $B_{a_3}\cup B_{a_4}$. Since the
graph was previously unlabeled, adding these new labels which do not appear in
$\phi$ does not affect the validity of $\phi$.  Note that this means that any
vertex of rank at least $i$ which appears in a bag in $[a_3,a_4]$ is now a
terminal.  Furthermore, any terminal of rank at least $i$ appears in all bags
of the section.  We define $t_1=a_3+5p\cdot\Delta(i-1)$ and
$t_2=a_4-5p\cdot\Delta(i-1)$.  We now have that all bags of the interval
$[t_1,t_2]$ contain the same set of terminals (because terminals of rank less
than $i$ appear in at most $5p\Delta(i-1)$ bags), therefore the terminals of
$B_{a_3},B_{a_4}$ of low rank do not appear in $[t_1,t_2]$.  This setup will
allow us to invoke Lemmas~\ref{lem:php1} and \ref{lem:php2} on the interval
$[t_1,t_2]$.

Recall that Lemmas~\ref{lem:php1} and \ref{lem:php2} use the values $L,
R=(L+1)(2^{(L+1)(p^2+kp+2p)}+1)$, and $R^*=(R+1)(q2^{(R+1)(p^2+kp+2p)}+1)$. We
will use similar values, but we will be a bit more generous with the constants
to allow us to more easily use the machinery of Section~\ref{sec:rewiring}.  We define
$\hat{L}=5p\cdot \Delta(i-1) \cdot(2^q-1)$,
$\hat{R}=(20\hat{L}+1)(2^{(20\hat{L}+1)(p^2+kp+2p)}+1)$, and
$\hat{R}^*=(5\hat{R}+1)(q2^{(5\hat{R}+1)(p^2+kp+2p)}+1)$.

We first show that $t_2-t_1>\hat{R}^*$. To see this, we will take some loose
upper bounds, simplifying things by using the fact that $x\le 2^x$ for any
non-negative integer $x$. We have $\hat{L} \le \Delta(i-1)\cdot p2^{q+3} \le
\Delta(i-1)\cdot 2^{p+q+3}$; $\hat{R} \le 2^{2(20\hat{L}+1)5p^2}\le
2^{250\hat{L}p^2}$, where we used that $k\le 2p$; $\hat{R}^*\le
2^{50\hat{R}qp^2}$. Putting these together we get that $\hat{R}^*\le
2^{50qp^2\cdot 2^{\left(250\hat{L}p^2\right)}}\le 2^{2^{300qp^2\hat{L}}}\le
2^{2^{300qp^2\cdot\left(\Delta(i-1)\cdot 2^{p+q+3}\right)}}\le
2^{2^{\Delta(i-1)\cdot 2^{10qp^2}}}$.  We now observe that $t_2-t_1\ge
\Delta(i)-10p\cdot\Delta(i-1)$, so it suffices to have that $\Delta(i) \ge
2^{2^{\Delta(i-1)\cdot 2^{10qp^2}}} + 10p\cdot\Delta(i-1)$. Since $\Delta(i) =
2^{2^{\Delta(i-1)\cdot 2^{20qp^2}}}$, this is clearly satisfied.

Because $t_2-t_1>\hat{R}^*$ and all bags of $[t_1,t_2]$ contain the same
terminals, we can invoke Lemma~\ref{lem:php2}, replacing $R$ with $5\hat{R}$.  We
get that there is a set of $q+1$ indices $S=\{s_1,\ldots,s_{q+1}\}$ in
$[t_1,t_2-5\hat{R}]$ such that for all $j,j'\in [q+1]$ we have
$|s_j-s_{j'}|>5\hat{R}$ and $s_j\approx_{5\hat{R}} s_{j'}$. Consider the set of
indices $S' = \{ s_j+2\hat{R}\ |\ j\in[q+1] \}$. It is not hard to see that for
any two elements $x,y\in S'$ we have $x-y>5\hat{R}$ and $x\approx_{\hat{R}} y$
(we are essentially looking at the middle fifth of each interval of length
$5\hat{R}$). We now apply Lemma~\ref{lem:foundid}, replacing $R$ with $\hat{R}$. We
conclude that if we apply the rewiring operations $q+1$ times, once in each
interval $[s,s+\hat{R}]$, for $s\in S'$, we will create $q+1$ identical new
parts in the graph as described in Lemma~\ref{lem:foundid}. What we now need to
argue is that we can apply this rewiring operation in appropriate places in
each such interval of length $\hat{R}$ so that the validity of $\phi$ is not
affected.

Consider now an interval $[s,s+\hat{R}]$, for $s\in S'$. We want to use
Lemma~\ref{lem:rewire} to show that applying the rewiring operation in two
appropriate bags of this interval does not affect the validity of $\phi$. Since
all such intervals are isomorphic for any $s\in S'$, it suffices to consider
one. We invoke Lemma~\ref{lem:php1} with $20\hat{L}$ in the place of $L$. We get
that there exist two indices $q_1,q_2\in [s,s+\hat{R}]$ such that $|q_2-q_1|\ge
20\hat{L}$ and $q_1\approx_{20\hat{L}} q_2$. Define $q_1'=q_1+10\hat{L}$,
$q_2'=q_2+10\hat{L}$. We have $|q_2'-q_1'|\ge 20\hat{L}$ and also that
$q_1,q_2$ are at distance at least $4\hat{L}$ from the endpoints of the
interval $[s,s+\hat{R}]$. Furthermore, $\mathcal{B}(q_1'-\hat{L},q_1'+\hat{L})$
is block-isomorphic to $\mathcal{B}(q_2'-\hat{L},q_2'+\hat{L})$. Since every
non-terminal appears in at most $5p\cdot\Delta(i-1)$ bags of the decomposition
and $\hat{L}=5p\cdot\Delta(i-1)\cdot(2^q-1)$, we can apply Lemma~\ref{lem:rewire}
and conclude that applying the rewiring operation on $q_1',q_2'$ does not
affect the validity of the formula.

We are now almost done. We have located $q+1$ pairs of indices such that
applying the rewiring operation on each pair does not affect the validity of
the formula, and such that the new parts we obtain from this operation are
identical. Invoking Lemma~\ref{lem:isomorphic}, we conclude that we can delete one
such part without affecting the validity of $\phi$. Let $G_1$ be the graph we
obtain when we do this. $G_1$ is indeed smaller than $G$ and we have $G\models
\phi$ if and only if $G_1\models \phi$. The only problem is that $G_1$ actually
has higher pathwidth than $G$, so we cannot simply return $G_1$.  To work
around this difficulty, we will ``undo'' the rewiring operation on the $q$
remaining index pairs.

More precisely, let $Q$ be the set of the $q+1$ pairs of indices where we
applied the rewiring operation and $(q_1,q_2)\in Q$ be one such pair.  Let
$G_2$ be the graph we obtain if instead of rewiring at $(q_1, q_2)$ we do the
following: we identify each vertex of $B_{q_1}$ with the vertex of $B_{q_2}$
which has the same rank; and we delete every vertex that appears in a bag of
$[q_1,q_2]$ without appearing in $B_{q_1}\cup B_{q_2}$. We now observe that
applying the rewiring operation on $G_2$ on every pair of $Q\setminus
\{(q_1,q_2)\}$ will produce a graph isomorphic to $G_1$. Indeed, what we did in
$G_2$ was delete the new part that would have been produced by the rewiring
operation on $(q_1,q_2)$ before rewiring the rest. But if this is the case,
with the same arguments as before we can invoke Lemma~\ref{lem:rewire} and claim
that $G_2\models \phi$ if and only if $G_1\models \phi$. Now, we are finally
done, because we can return $G_2$. A path decomposition of $G_2$ can easily be
obtained from the path decomposition of $G$ and we retain the same ranking
function and drop the labeling function, which we no longer need.  \end{proof}

\begin{thm}\label{thm:elem} For every fixed $p$, model checking a formula $\phi$ on a graph
$G$ with pathwidth $p$ can be performed in time $f(\phi) |G|^{O(1)}$, where $f$
is an elementary function.  \end{thm}

\begin{proof} Let $q$ be the number of quantifiers in $\phi$. We first compute
a normalized path decomposition using Lemma~\ref{lem:normal}. Let $p'\le 8p$ be the
number of ranks used by the ranking function.

We first check if the graph contains a vertex that appears in more than
$5p'\Delta(p')$ bags, on which we could apply Lemma~\ref{lem:main}.  Clearly, this
can be done in time $\Delta(p') n^{O(1)}$, which is FPT with an elementary
dependence on $q$ (if $p$ is a fixed constant).  If we find such a vertex, we
invoke Lemma~\ref{lem:main} and obtain a smaller graph together with a path
decomposition of the same width. We repeat this process as long as possible,
and this adds at most a factor of $n$ to the running time.

Once we can no longer find such a vertex, we have an elementary (in $q$) upper
bound on the maximum degree of the current graph, because a vertex that appears
in $\delta$ bags can have at most $p\delta$ neighbors.  We then invoke the
algorithm of \cite{Seese96} which handles FO model checking on graphs of
bounded degree and has an elementary dependence on $q$, and output the result
of this algorithm.  \end{proof}

\section{Conclusions}

We have shown that FO model checking for graphs of bounded pathwidth has a
complexity behavior that is in sharp contrast with both MSO logic for the same
class of graphs and the complexity of FO logic on graphs of bounded treewidth.
One way in which it may be interesting to improve upon our result is to note
that our algorithm's dependence on the pathwidth is a tower of exponentials
whose height is $O(\pw)$, where the hidden constant is roughly $16$.  This is
in contrast with the meta-theorem of \cite{Gajarsky15}, where the height of the
tower is roughly equal to the tree-depth. It would be interesting to
investigate if the height of the tower in our case can be made $\pw+O(1)$, or
conversely to prove that this is not possible and FO model checking on bounded
pathwidth is truly harder than for bounded tree-depth. We note that the current
constant ($16$) is affected by the state of the art on the performance of the
First-Fit algorithm on interval graphs \cite{NB08}, which is $8$-competitive,
so one way to improve the algorithm would be to replace this argument. One
possible approach to this problem (suggested by an anonymous reviewer) would be
to use Simon's factorization theorem.

\bibliographystyle{alphaurl}
\bibliography{fopathwidth}

\end{document}